\documentclass[tradiabstract]{aa}  

\usepackage{graphicx}
\usepackage[dvipsnames]{xcolor}
\usepackage{txfonts}
\usepackage{appendix}
\usepackage{hyperref}
\usepackage{soul}  
\usepackage{xparse}  
\usepackage{comment}
\usepackage{amssymb}

\begin{document} 

\newcommand{\Ho}{\mbox{$H_0$}}
\newcommand{\ang}{\mbox{{\rm \AA}}}
\newcommand{\abs}[1]{\left| #1 \right|} 
\newcommand{\kms}{\ensuremath{{\rm km\,s^{-1}}}}
\newcommand{\zabs}{\ensuremath{z_{\rm abs}}}
\newcommand{\zem}{\ensuremath{z_{\rm quasar}}}
\newcommand{\cmsq}{\ensuremath{{\rm cm}^{-2}}}
\newcommand{\ergs}{\ensuremath{{\rm erg\,s^{-1}}}}
\newcommand{\ergsa}{\ensuremath{{\rm erg\,s^{-1}\,{\AA}^{-1}}}}
\newcommand{\ergscm}{\ensuremath{{\rm erg\,s^{-1}\,cm^{-2}}}}
\newcommand{\ergscma}{\ensuremath{{\rm erg\,s^{-1}\,cm^{-2}\,{\AA}^{-1}}}}
\newcommand{\msyr}{\ensuremath{{\rm M_{\rm \odot}\,yr^{-1}}}}
\newcommand{\nhi}{n_{\rm HI}}
\newcommand{\fhi}{\ensuremath{f_{\rm HI}(N,\chi)}}
\newcommand{\refs}{{\bf (refs!)}}
\newcommand{\Av}{\ensuremath{A_V}}
\newcommand{\lya}{Ly-$\alpha$}
\newcommand{\Hb}{H-$\beta$}
\newcommand{\OVI}{\ion{O}{vi}}
\newcommand{\OIII}{\ion{O}{iii}}
\newcommand{\OII}{\ion{O}{ii}}
\newcommand{\OI}{\ion{O}{i}}
\newcommand{\HI}{\ion{H}{i}}
\newcommand{\HeII}{\ion{He}{ii}}
\newcommand{\HH}{\ensuremath{{\rm H}_2}}
\newcommand{\SII}{\ion{S}{ii}}
\newcommand{\SiIII}{\ion{Si}{iii}}
\newcommand{\SiIV}{\ion{Si}{iv}}
\newcommand{\SiII}{\ion{Si}{ii}}
\newcommand{\AlIII}{\ion{Al}{iii}}
\newcommand{\AlII}{\ion{Al}{ii}}
\newcommand{\ArI}{\ion{Ar}{i}}
\newcommand{\FeII}{\ion{Fe}{ii}}
\newcommand{\ZnII}{\ion{Zn}{ii}}
\newcommand{\CrII}{\ion{Cr}{ii}}
\newcommand{\MnII}{\ion{Mn}{ii}}
\newcommand{\MgII}{\ion{Mg}{ii}}
\newcommand{\MgI}{\ion{Mg}{i}}
\newcommand{\NiII}{\ion{Ni}{ii}}
\newcommand{\NV}{\ion{N}{v}}
\newcommand{\CIV}{\ion{C}{iv}}
\newcommand{\CIII}{\ion{C}{iii}}
\newcommand{\CII}{\ion{C}{ii}}
\newcommand{\CI}{\ion{C}{i}}
\newcommand{\CaII}{\ion{Ca}{ii}}
\newcommand{\TiII}{\ion{Ti}{ii}}
\newcommand{\Jzzuc}{J0015$+$1842}
\newcommand{\Jzzun}{J0019$-$0137}
\newcommand{\Jzzcn}{J0059$+$1124}
\newcommand{\Jzudc}{J0125$-$0129}
\newcommand{\Jzuts}{J0136$+$0440}
\newcommand{\Jzhch}{J0858$+$1749}
\newcommand{\Judts}{J1236$+$0010}
\newcommand{\Judqh}{J1248$+$0639}
\newcommand{\Judcn}{J1259$+$0309}
\newcommand{\Juttu}{J1331$+$0206}
\newcommand{\Jutch}{J1358$+$1410}
\newcommand{\Jdddh}{J2228$-$0221}
\newcommand{\Jdtdc}{J2325$+$1539}
\newcommand{\si}{$S_{\rm ion}$}
\newcommand{\sn}{$S_{\rm ion + neu}$}

\newcommand{\fcla}{French-Chilean Laboratory for Astronomy, IRL 3386, CNRS and U. de Chile, Casilla 36-D, Santiago, Chile \label{fcla}}
\newcommand{\iap}{Institut d'Astrophysique de Paris, CNRS-SU, UMR\,7095, 98bis bd Arago, 75014 Paris, France -- \email{noterdaeme@iap.fr}\label{iap}}
\newcommand{\ioffe}{Ioffe Institute, {Polyteknicheskaya 26}, 194021 Saint-Petersburg, Russia \label{ioffe}}
\newcommand{\cral}{Centre de Recherche Astrophysique de Lyon, UMR5574, U. Lyon 1, ENS de Lyon, CNRS, 69230 Saint-Genis-Laval, France \label{cral}}
\newcommand{\uchile}{Departamento de Astronom\'ia, Universidad de Chile, Casilla 36-D, Santiago, Chile \label{uchile}}
\newcommand{\oat}{INAF - Osservatorio Astronomico di Trieste, Via G.B. Tiepolo, 11, I-34143 Trieste, Italy \label{oat}}

\definecolor{green}{rgb}{0,0.4,0}

\DeclareDocumentCommand{\PN}{s m O{}}{{\color{red}\IfBooleanTF{#1}{\color{gray} \setstcolor{red}\st{#2} \color{red}#3}{{\sl [PN: #2]}}}}
\DeclareDocumentCommand{\SB}{s m O{}}{{\color{violet}\IfBooleanTF{#1}{\st{#2} #3}{{\sl [SB: #2]}}}}
\newcommand{\RC}[1]{{\color{blue} RC:~ #1}}


\newcommand{\old}[2]{{\color[rgb]{0,0,0}\sout{#1}}{\color[rgb]{0.7,0,0.0}{\bf #2}}}
\newcommand{\new}[1]{{\bf #1}}
\newcommand{\assign}[2]{\noindent{\color{blue}#1 $\to$ #2}}

\defcitealias{Noterdaeme2019}{N19} 	 
\defcitealias{Noterdaeme2023}{N23}

\title{Exploring quasar evolution with proximate molecular absorbers: Insights from the kinematics of highly ionized nitrogen\thanks{Based on observations collected at the Very Large Telescope of the European Southern Observatory, Prgm IDs 094.A-0362, 103.B-0260 and 105.203L.001.}}
\titlerunning{\NV\ in quasars with proximate H$_2$}

 \author{
   R.~Cuellar\inst{\ref{uchile}, \ref{fcla}}
   \and
   P.~Noterdaeme\inst{\ref{fcla},\ref{iap}}
  \and
   S.~Balashev\inst{\ref{ioffe}}
   \and
   S.~L{\'opez}\inst{\ref{uchile}} 
\and
   V. D'Odorico\inst{\ref{oat}}
   \and 
   J.-K.~Krogager\inst{\ref{fcla},\ref{cral}}
   }
   
   \institute{\uchile \and  \fcla \and \iap \and \ioffe \and \oat \and \cral}
             
   \date{\today.}

    \abstract{
We investigate the presence and kinematics of \NV\ absorption proximate to high redshift quasars selected upon the presence of both strong H$_2$ and \HI\ absorption at the quasar redshift. 

Our spectroscopic observations with X-shooter at the VLT reveal a 70\% detection rate of \NV\ (in 9 out of 13 quasars with $2.5 < z < 3.3$), remarkably higher than the $\sim$10\% detection rate in intervening Damped Lyman-$\alpha$ systems and the $\sim$30\% rate observed within a few thousand \kms\ of the source in the general quasar population.

While many \NV\ components lie within the velocity range of the neutral gas, the kinematic profiles of high-ionization species appear decoupled from those of low-ionization species, with the former extending over much larger velocity ranges, particularly towards bluer velocities (up to several thousand \kms). We also observe significant variations in the \NV\ to \SiIV\ ratio, which we attribute to varying ionization conditions, with a clear velocity-dependent trend: blueshifted \NV\ components systematically exhibit higher ionization parameters compared to those near the quasar's systemic redshift. Furthermore, the most redshifted systems relative to the quasar show no evidence of \NV\ absorption.

The results suggest that proximate H$_2$ absorption systems select critical stages of quasar evolution, during which the quasar remains embedded in a rich molecular environment. Redshifted systems likely trace infalling gas, potentially associated with mergers, preceding the onset of outflows. Such outflows, as traced by \NV, may eventually reach or even carry out neutral and molecular gas. This latter stage would correspond to proximate H$_2$ systems located around or blueshifted relative to the quasar's systemic redshift.

Finally, the only case in our sample featuring highly blueshifted neutral gas (-2000~\kms) shows no evidence of an association with the quasar. Our findings highlight the need to account for the ionization state when defining a velocity threshold to distinguish quasar-associated systems from intervening ones.
}


\maketitle
\section{Introduction \label{s:Introduction}}

Understanding the co-evolution of active galactic nuclei (AGN) and their host galaxies is a major topic in astronomy \citep{Volonteri2021,Morganti2017}. While the presence of the bright nucleus bothers observations of host and companion galaxies in emission, 
it becomes an advantage to study the gas in absorption. For example, the presence of neutral gas along the line of sight with $N(\HI) \ge 2\times 10^{20}$\,cm$^{-2}$ is revealed by the so-called Damped Lyman-$\alpha$ systems \citep[DLAs, see e.g.][]{Wolfe2005}.

DLAs can be classified into two primary categories: intervening and associated, depending on their origin with respect to the background sources. Intervening DLAs are fortuitous encounters  that are not related with the sources themselves. The incidence, column density and metallicity distribution of intervening DLAs indicate a close connection of them with the overall population of galaxies \citep[e.g.][]{Prochaska2005, Pontzen2008, Krogager2017, Krogager2020}. 

When the redshift of an absorption system lies within a few thousand \kms\ of the source's emission redshift, it is better referred to as a 'proximate' \citep[][]{Ellison2002,Ellison2010,Prochaska2008PDLA}. While the proximity in velocity space is immediate, it remains non-trivial to determine whether a given proximate DLA (PDLA) originates in the AGN host galaxy, its environment, or if it has any association with the source at all \citep[e.g.][]{Moeller1998}\footnote{
For these reasons, we favor "proximate" over "associated" \citep[e.g.][]{Weymann1979}, traditionally defined by relative velocity alone, until physical association is confirmed.}.
Statistically, the higher incidence of DLAs at small velocity separations from quasars aligns with the expected galaxy overdensity, despite the counteracting effect of the quasar’s intense radiation, which ionizes gas over large distances \citep[the so-called proximity effect]{Prochaska2008}.

There is also evidence from metal lines suggesting that PDLAs exhibit distinct characteristics compared to intervening systems \citep{Fechner2009,Ellison2010,Ellison2011}.
\citet{Finley2013} also noted the presence of strong Ly-$\alpha$ emission in the saturated core of at least a fraction of proximate DLAs. The strength and width of this emission are most likely explained by the Ly-$\alpha$ emission from the quasar not being fully covered by the absorber as opposed to much weaker emission due to in-situ star-formation as directly detected in a few intervening DLAs \citep[e.g.][]{Moller2004,Fynbo2010,Noterdaeme2012,Krogager2013,Ranjan2018} or statistically from stacking fiber spectra \citep[e.g.][]{Rahmani2010,Noterdaeme2014,Dharmender2024}. Interestingly, this also revealed a potential bias against the identification of PDLAs since the most recognisable characteristic of DLAs is normally their zero flux level over a wide velocity range. 
A population of PDLAs where the damped core is almost completely filled with emission has indeed been identified,
with a primary identification based on metal absorption lines \citep{Fathivavsari2018, Fathivavsari2020}. 

More recently, \citet[hereafter \citetalias{Noterdaeme2019}]{Noterdaeme2019} discovered a population proximate molecular-rich DLAs detected solely based on the Lyman-Werner absorption bands of H$_2$ in SDSS quasar spectra. Not only the incidence of proximate H$_2$  
absorbers is in strong excess to what is expected from intervening statistics, but H$_2$ is a sensitive tool to investigate the physical conditions in the gas \citep[e.g.][]{Balashev2019,Balashev2020,Klimenko2020,Kosenko2024}. For example, it has been possible to use H$_2$ together with excited atomic species to measure the distance between the absorbing gas and the quasar and to reveal an origin in a multi-phase outflow \citep[]{Noterdaeme2021}. \citet[\citetalias{Noterdaeme2023}]{Noterdaeme2023} recently presented the follow-up of a sample of 13 proximate H$_2$ systems observed with X-shooter on the VLT. The measurement of kinematics aided by precise measurements of the quasar's systemic redshift (from NIR lines and/or CO emission lines with NOEMA) together with a study of chemical abundances strengthen an origin of these systems in the environment of the quasar. 

Remarkably, \NV\ absorption (i.e. four times ionized nitrogen, N$^{4+}$) appears to be frequent in these PDLAs. This is particularly interesting since ionizing nitrogen to such level requires an energy of 77.5~eV, not produced by stellar light (which falls off rapidly after the \HeII\ ionization energy at 54~eV).
Indeed \NV\ is rarely seen in intervening DLAs but seems to be {more common in metal absorption systems within 5000 \kms\ around quasars} \citep{Perrotta2016,Perrotta2018}, and hence a likely indication of physical proximity as well. 

In this paper, we investigate the presence of \NV\ absorption more in depth thanks to velocity decomposition of the profiles and the observation of associated \SiIV. 
We present our sample in Sect. ~\ref{s:Data} and the measurements in Sect.~\ref{s:measurements}. We perform photo-ionization modelling to aid the interpretation of the observations in Sect. ~\ref{s:cloudy_models}. 
The results are presented and discussed in Sect.~\ref{s:Results} and we 
conclude in Sect. ~\ref{s:conclusion}. We assume a flat $\Lambda$CDM cosmology with H$_{O}$ = 68 \kms\,Mpc$^{-1}$, $\Omega_{\Lambda}$ = 0.69, and $\Omega_{m}$ = 0.31 \citep{PlanckCollaboration2016}. 
The column densities, denoted $N$, are provided in units of cm$^{-2}$.

\section{Sample and detection statistics\label{s:Data}}

Our sample consists of 13 high-redshift (\textbf{$2.5 < z < 3.3$}) proximate H$_2$ systems from \citetalias{Noterdaeme2019}, observed at {resolving} power $R \sim 6000$–10\,000 with the ESO-VLT X-shooter spectrograph \citep{Vernet2011}. {The spectra were processed using the offical esorex pipeline version 3.5.3 \citep{Modigliani2010}, and {line profiles were modeled} through a standard multi-component Voigt-profile fitting.}
We refer to \citetalias{Noterdaeme2023} for comprehensive details on the selection criteria, observations, data reduction, and measurements of column densities for low-ionization ions, from which gas-phase metallicities were derived. The basic information is {presented} in Table~\ref{t:sum}.

Here, we focus on highly ionized species, specifically \NV\ and \SiIV, and to some extent, \CIV. These ions produce doublet lines conveniently located redwards of the \lya\ emission in proximate systems, simplifying their detection and measurement.
We clearly detected \NV\ doublets in nine out of thirteen systems in our sample, yielding a $\sim$70\% detection rate --significantly higher than the rates observed so far in intervening DLAs ($\sim$13\% reported by \citealt{Fox2009}), with similar detection limits ($\log N(\NV)\sim$ 13).

Using high-resolution spectra of nearly a hundred DLAs, \citet{Fox2009} reported a similar \NV\ detection rate of 13\% for both intervening systems (10/75) and proximate systems (2/16, within 5000~\kms\ of the quasar). This similarity, which contrasts with the expectation of an enhanced \NV\ detection rate due to the hard radiation field near quasars, may indicate that the quasar's radiation field actually dominates over a smaller velocity range than previously assumed: 
At a few thousand \kms, most proximate DLAs in the literature may still be intervening systems rather than being located in the quasar's immediate environment. For instance, at $z \sim 2-3$, 3000~\kms\ in the Hubble flow corresponds to a distance of approximately 10~Mpc, where the quasar’s UV radiation field diminishes to the level of the metagalactic background.

Here, in spite of their original detection in low-resolution spectra \citepalias{Noterdaeme2019}, which in principle allowed for large velocity shifts from the quasar, all but one\footnote{As we discuss later, the system with the largest velocity offset in our sample, at $v \sim -2000$~\kms\ relative to quasar \Jzhch, lacks evidence of physical proximity and may instead represent an intervening system \citep[see also][]{Balashev2019}.} H$_2$-selected system in our sample lie within 1000~\kms\ of the quasar's systemic redshift \citepalias{Noterdaeme2023}, suggesting a stronger physical association with the quasar than seen in regular PDLAs \citep[as in the sample by][]{Fox2009}, which could explain our higher \NV-detection rate. 

Interestingly, {we found that} \NV\ absorption can also appear at large velocities relative to neutral gas. Among the nine systems with \NV\ detections, one shows components confined to the velocity range of low-ionization species, seven exhibit components both within and beyond this range, and one contains components exclusively at high velocity separations.

\label{sec:coldens}

\begin{table}[]
\caption{Quasar sample and \NV-detection summary. \label{t:sum}}

\resizebox{\hsize}{!}{
\begin{tabular}{cccccc}
\hline  \hline
{\large \strut}Quasar & RA (J2000) &  DEC (J2000) & $z_{\mathrm{sys}}$\tablefootmark{a}   & $z_{{50}}$\tablefootmark{b} & \NV\ \\
\hline 
\Jzzuc &  00:15:14.82  &  18:42:12.34   & 2.6310  & 2.6288 & yes \\        
\Jzzun &  00:19:30.55  &  -01:37:08.46  & 2.5217  & 2.5281 & yes \\        
\Jzzcn &  00:59:17.64  &  11:24:07.70   & 3.0369  & 3.0342 & yes \\        
\Jzudc &  01:25:55.11  &  -01:29:25.00  & 2.6534  & 2.6636 & no \\        
\Jzuts &  01:36:44.02  &  04:40:39.10   & 2.7849  & 2.7796 & yes \\        
\Jzhch &  08:58:59.67  &  17:49:25.19   & 2.6478  & 2.6254 & no \\        
\Judts &  12:36:02.11  &  00:10:24.54   & 3.0284  & 3.0328 & yes \\        
\Judqh &  12:48:29.51  &  06:39:35.62   & 2.5273  & 2.5292 & yes \\        
\Judcn &  12:59:17.31  &  03:09:22.51   & 3.2365  & 3.2461 & no \\        
\Juttu &  13:31:11.41  &  02:06:09.06   & 2.9147  & 2.9220 & yes \\        
\Jutch &  13:58:08.94  &  14:10:53.29   & 2.8926  & 2.8925 & yes \\        
\Jdddh &  22:28:07.36  &  -02:21:17.17  & 2.7705  & 2.7690 & yes \\        
\Jdtdc &  23:25:06.62  &  15:39:29.31   & 2.6072  & 2.6167 & no \\        
\hline
\end{tabular}}
\tablefoot{\tablefoottext{a}{Systemic redshift derived from the quasar emission lines \citepalias[see][]{Noterdaeme2023}.}
\tablefoottext{b}{Centroid of the neutral gas defined as the redshift where the cumulative optical depth of low-ionization species reaches 50\% of its maximum value.}}
\end{table}

\section{Data analysis\label{s:measurements}}
A main assert of the present work is the possibility to resolve the line profiles and hence to investigate how physical quantities vary along the kinematical profile, and in particular, whether high-ions components can have some degree of association with the neutral gas (at least probing the same object) or arise from a distinct phenomena. While coinciding absorption components across species do not guarantee co-spatiality, differing velocities between species indicate separate clouds, though they may still trace the same overarching structure.

We performed simultaneous multi-component Voigt-profile fitting of the detected \NV$\lambda\lambda$1238,1242, \SiIV$\lambda\lambda$1393,1402 and  {\CIV$\lambda\lambda$1548,1550} doublets using the {\sc Voigtfit} code \citep{krogager2018voigtfit}. This approach enabled us to determine the relative velocities, Doppler parameters ($b$), and column densities ($N$) for each component within the systems.
Because \CIV\ lines are strongly saturated in most cases, the derived column densities are generally highly uncertain. However, \CIV\ remains useful to constrain the velocity structure and ionization, particularly in weaker components where \NV\ is detected but \SiIV\ is below the detection threshold.
The Voigt-profile fits are shown in Appendix \ref{s:indv_spectra} and the column densities in \NV-bearing components provided in Table~\ref{t:column_density}.

\begin{figure}
\centering
    \includegraphics[width = \hsize,trim={0 0.5cm 1.cm 1cm},clip]{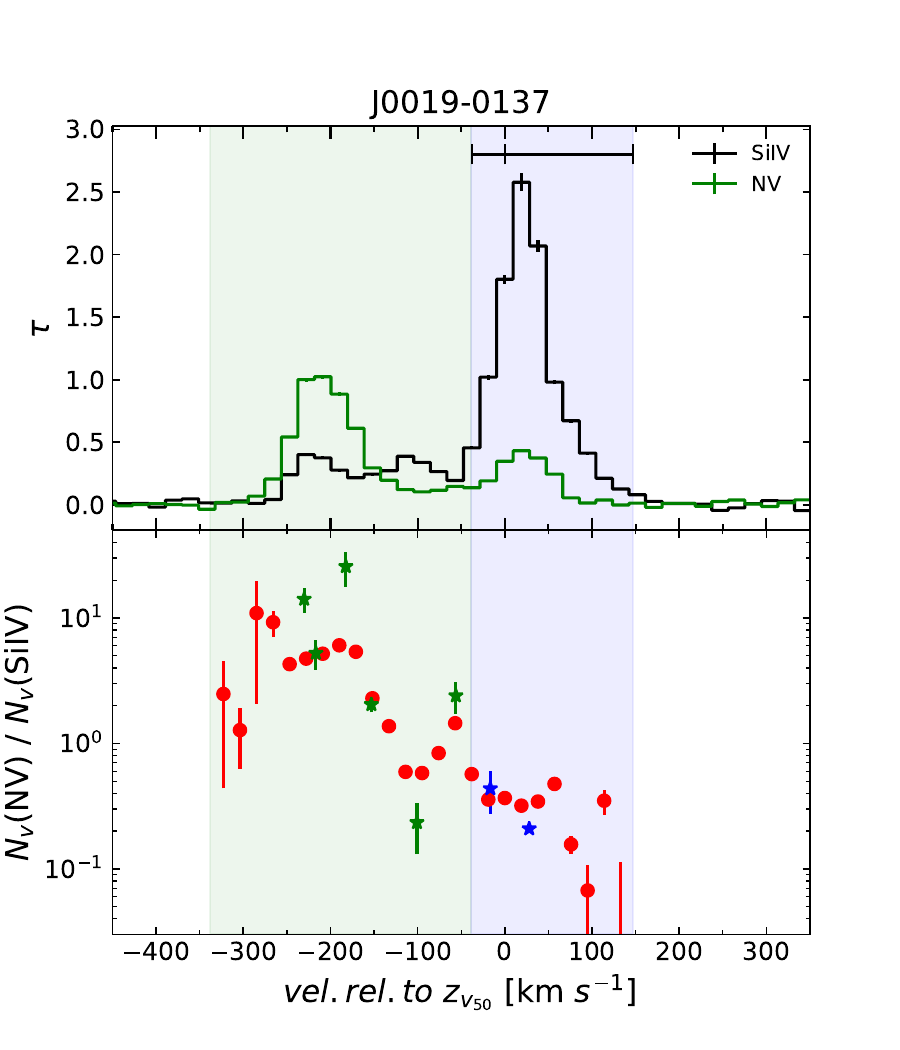}
    \caption{Top: Apparent optical depth of the \NV$\lambda\lambda$1338 (green) and \SiIV$\lambda\lambda$1393 (black) absorption lines towards \Jzzun. The velocity extent of the low-ionization species ($\Delta v_{\rm 90}$) is marked by the blue shaded region and the horizontal segment, with the tick marks corresponding to the 5, 50 and 95 percentiles of the cumulative optical depth \citep[see][]{Prochaska1997}. Bottom: Apparent column density ratio per unit velocity bin (red dots) as well as ratio obtained for individual components used in the Voigt-profile modelling (stars). 
    \label{f:AOD}}
\end{figure}

In Fig.~\ref{f:AOD}, we compare the \NV/\SiIV\ column density ratio derived from Voigt-profile fitting with that obtained pixel-by-pixel using the apparent optical depth method \citep[][]{Savage1991} for the quasar \Jzzun. Both methods yield consistent results and reveal significant variations in the \NV/\SiIV\ ratio across the profile. Notably, there is a tendency for higher ratios at velocities outside the bulk of the neutral gas, quantified by $\Delta v_{90}$—the velocity range encompassing 90\% of the total optical depth of low-ionization species \citepalias[from][]{Noterdaeme2023}.
While the apparent optical depth method offers simplicity and uniqueness, it is sensitive to the signal-to-noise ratio and saturation effects \citep{Fox2005}, and it cannot effectively account for line blending. Since blending is evident in several other cases, we use the Voigt-profile-derived quantities for subsequent analysis, keeping in mind that the decomposition may not be unique, as in most absorption studies.

\section{Photo-ionization models  \label{s:cloudy_models}}

In order to understand the physical origin of the variation in the \NV-to-\SiIV\ ratio, we perform photo-ionization models with {Cloudy c.23} \citep[last described in][]{Chatzikos2023}, assuming a plane-parallel geometry of gas under a hard UV radiation field, likely dominated by the quasar. 

The incident radiation field is thus composed of the quasar radiation {("AGN field" in Cloudy)}, 
plus the metagalactic background taken from \citet[][KS19]{Khaire2019} at the redshift of the system. We checked that varying the adopted AGN field shape \citep[following][]{Vasudevan2009,Lusso2010,Melendez2011,DelMoro2017}, does not significantly affect the predicted column densities.

\begin{figure}
    \includegraphics[width =\hsize,trim={0.05cm 0 1.3cm 0.8cm},clip]{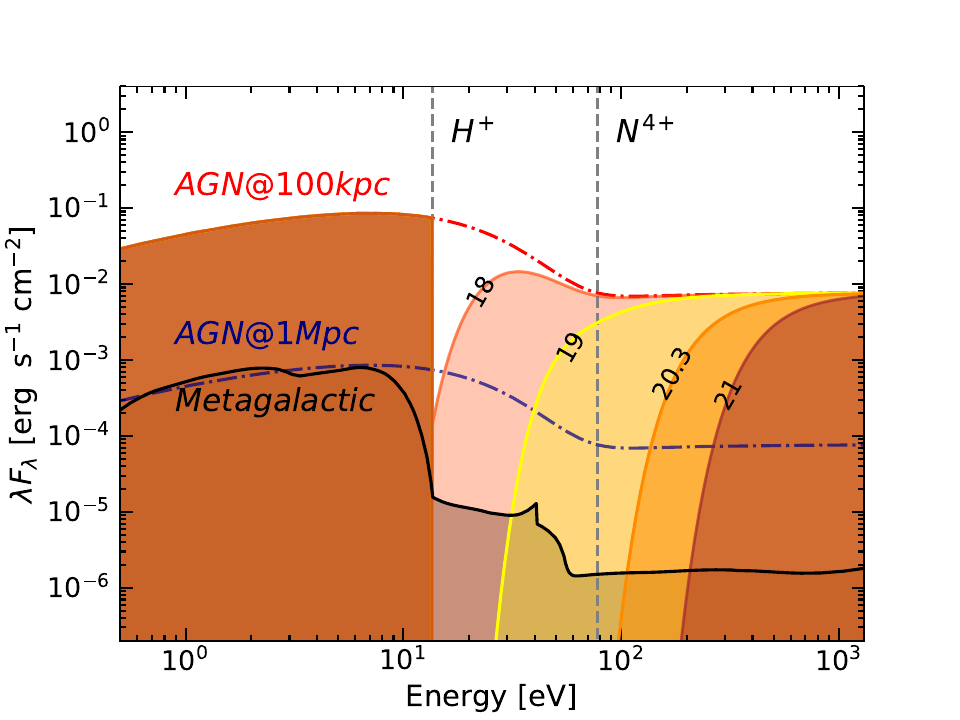}
    \caption{Comparison of incident radiation fields. The metagalactic background (KS19), {at redshift 2.6}, is represented by the black line. The red (blue) dashed line represents the unattenuated AGN field, with typical luminosity of our sample, at a distance of 100 kpc (1~Mpc). The other lines depicts the transmitted flux through a \HI\ layer with logarithmic column density values of 18 (pink), 19 (yellow), 20.3 (orange), and 21 (brown). The vertical dashed lines correspond to the ionization energy of hydrogen and that required to ionize nitrogen to N$^{4+}$.}
    \label{f:incident_flux}
\end{figure}

The incident flux in the model is represented in Fig.~\ref{f:incident_flux} over the relevant energy range for the present study. For the typical quasar luminosity ({ $L_{1450}$ =  10$^{45.8}$ erg $s^{-1}$ $\AA^{-1}$}) in our sample, the incident UV field is dominated by the AGN up to about 1~Mpc, where it becomes comparable to the metagalactic background.

Since we do observe strong damped Lyman-$\alpha$ absorption {in our sample}, we also show the effect of a \HI\  layer on the transmitted radiation field, which becomes zero at 13.6~eV and only slowly recovers at higher energies.
In the presence of neutral hydrogen, the number of available photons at energies above 77.5~eV --those capable of ionizing nitrogen to N$^{4+}$-- is drastically reduced at the DLA column density threshold. This indicates that the ionized gas containing \NV\ is likely located upstream, closer to the quasar than the bulk of the neutral (\HI-bearing) gas.

Since we have no direct measurement of the metallicity in the ionized phase, we use the metallicity from volatile species observed in the neutral phase {\citepalias[from][]{Noterdaeme2023}} and assume solar relative abundances \citep{Grevesse2010}.
We note that a high-ionization phase might carry the bulk of the metals in an outflow. As a result, the low- and high-ionization phases may exhibit different metallicities. The assumed metallicity has however very little effect on the predicted column density ratios, since it results in an almost linear scaling of the column densities (Fig.~\ref{f:N_and_ratio_N_vs_U_dif_Z}). As we will show later, this does not impact the derived ionization parameters.
Because we focus here on ionized gas under the radiation field of the quasar, in which dust grains should be easily destroyed, we neglect dust-depletion.

\begin{figure}
    \centering
    \includegraphics[width = 0.9\hsize,trim={0.2cm 0 1.5cm 0.8cm},clip]{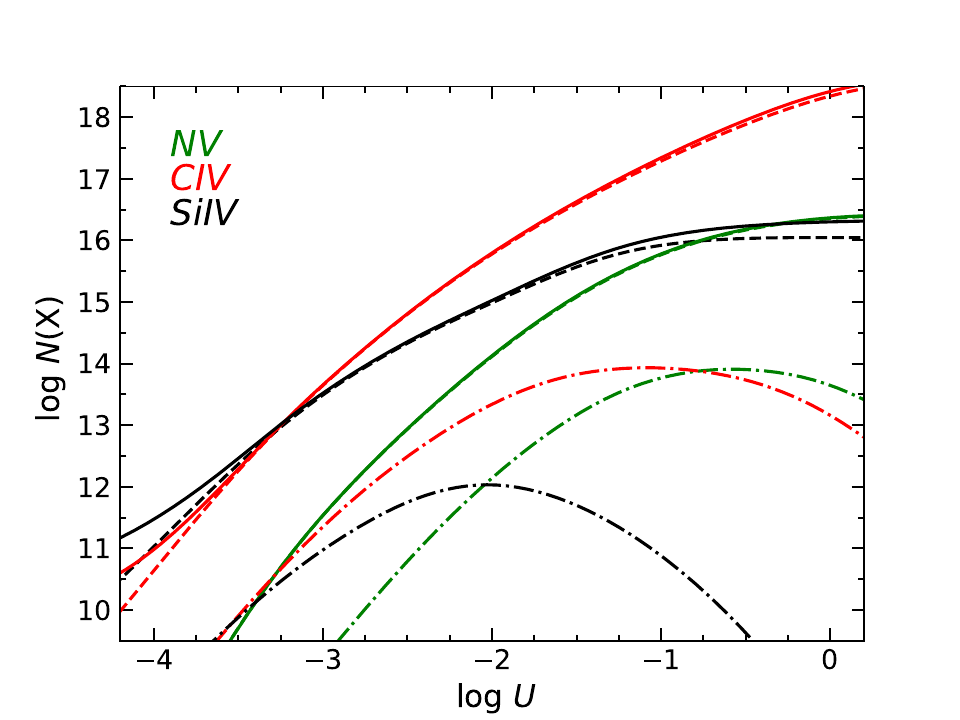}

    \caption{{Modeled} column density of high ionization species (colour-coded) as a function of the ionization parameter for different total \HI\ column density used as stopping criteria for the calculations. 
    The models were here ran with $\log Z/Z_{\odot}=-0.9$ (median metallicity in the sample) 
    and $\log n_{\rm H}$/${\rm cm}^{-2} = 2$. 
    The dashed-dotted, dashed and solid correspond to $\log N(\HI)=15$, 18 and 21, respectively.
} 
    \label{f:N_vs_U}
\end{figure}

An important unknown is the hydrogen column density of the clouds, to which the column densities of the metal species are directly scaled. For example, \citet{Perrotta2018} used a somehow arbitrary $\log N({\rm H}) = 20$ as a stopping criterion when modeling the total integrated metal column densities. 
Here, since we have a measurement of the total \HI\ column densities, {one could be tempted to use the observed $N(\HI)$ as a stopping criterion}. However, this approach would imply treating the full system as a single cloud, {while in reality it} certainly includes a mix of clouds and phases, as evidenced by the presence of multiple velocity components --in which we cannot determine the individual \HI\ column density-- and significant variations in observed \NV/\SiIV\ ratio between these components. 
Furthermore, the strong reduction of ionizing photons impedes the production of highly ionized species in the neutral gas, so that pushing the calculation to that depth would not bring any additional information on the ionized gas. 

This is further illustrated in Fig.~\ref{f:N_vs_U}, which presents the calculated column densities of various species as a function of the ionization parameter $U \equiv n_{\gamma} /n_{\rm H}$, where $n_{\gamma}$ is the number density of incident ionizing photons and $n_{\rm H}$ the number density of hydrogen atoms, for different \HI\ column densities and assuming the mean metallicity in our sample, $\log Z/Z_{\rm \odot}=-0.9$. 
As expected, the results concerning the highly ionized species are almost indistinguishable for $\log N(\HI)=18$ and 21, i.e. despite continuing the calculation up to three orders of magnitude higher \HI\ column densities, no significant amounts of highly ionized metals are produced once entering the \HI-dominated regime.
In contrast, the predicted column densities of \SiIV\ , \CIV\ and \NV\ vary strongly between $\log N(\HI)=15$ and 18. 
Solely from this calculation we observe that the detection of \SiIV\ at the observed levels must imply a column density of $\log N(\HI)>15$, since otherwise the predicted values do not exceed $\log N(\SiIV)=12$ regardless of the ionizing parameter, contingent upon the assumed metallicity (here $\log Z/Z_{\rm \odot}=-0.9$). On the other hand, the detection of \NV\ with column densities $\log N(\NV) \sim $ 12 - 16 implies ionization parameters likely above $\log U \sim -2.5$.

\begin{figure}
    \includegraphics[trim={0 0 1.45cm 0.5cm},clip,width = \hsize]{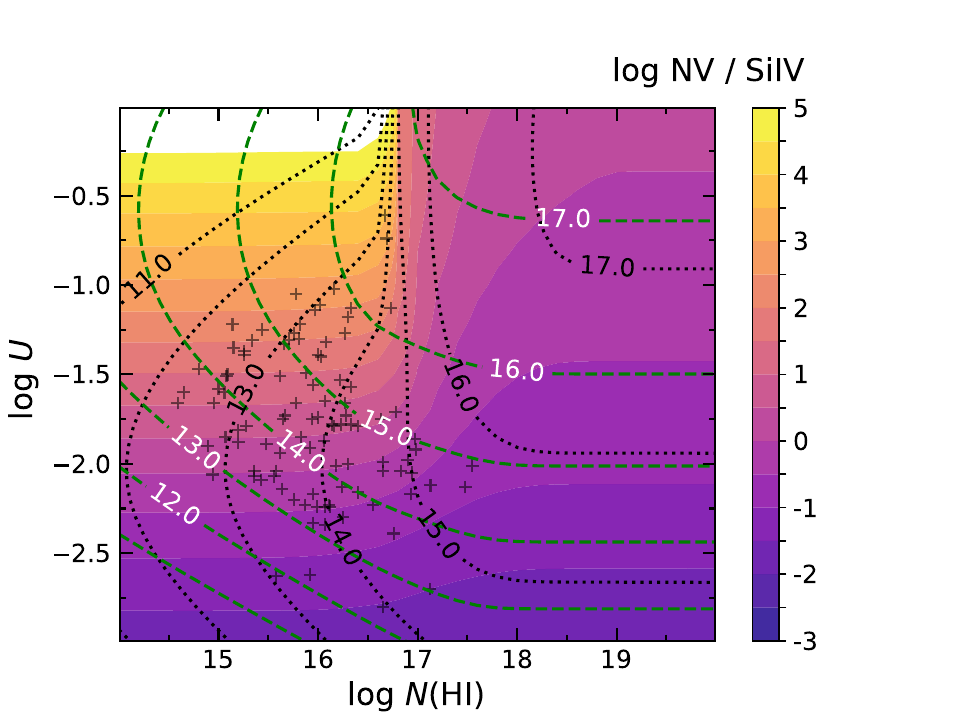}
    \caption{
    {\NV/\SiIV} ratio (colour scale) as a function of the ionization parameter and the \HI\ column density in the cloud. The green dashed and black dotted lines represent the column densities of \NV\ and \SiIV\ scaled by the metallicity, i.e.  $\log N(\NV)-\log Z/Z_{\rm \odot}$ and $\log N(\SiIV)-\log Z/Z_{\rm \odot}$. Crosses correspond to {measurements in the individual components in our sample.} Typical uncertainties on the observed column densities are about 0.2~dex. While these are considered when inferring the $U$ and $N(\HI)$ ranges, they are not represented here to avoid overcrowding the figure.
    }      
    \label{f:maps_logU_NHI}
\end{figure}

In Fig. ~\ref{f:maps_logU_NHI}, we illustrate how the \NV\ and \SiIV\ column densities depend jointly on the ionization parameter and the \HI\ column density. The metallicity is simply treated as direct scaling factor (see Fig.~\ref{f:N_and_ratio_N_vs_U_dif_Z}). 
For values either below $\log N(\mbox{\HI})\sim17$ or above $\log N(\mbox{\HI})\sim18$, the \NV\ to \SiIV\ ratio depends then primarily on the ionization parameter while being insensitive to the exact \HI\ column density. 

\begin{figure}
    \includegraphics[trim={0 0 1.45cm 0.5cm},clip,width = \hsize]{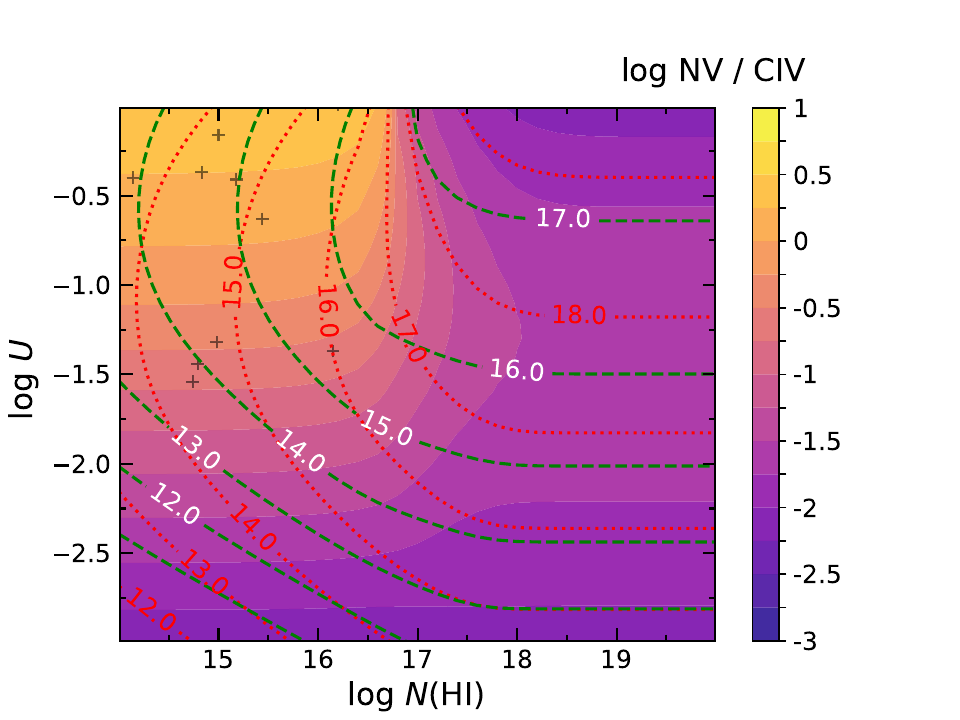}
    \caption{Same as Fig.~\ref{f:maps_logU_NHI} using \NV\ (green-dashed) and \CIV\ (red-dotted). 
    \label{f:maps_logU_NHI_CIV}}
\end{figure}

The dependence on ionization parameter is however different in the low ($\log N(\HI)<17$) and high ($\log N(\HI)>18$) regimes. In the intermediate regime, i.e. crossing the H-ionization front, \NV/\SiIV\ depends strongly on both $N(\HI)$ and $U$. In Fig.~\ref{f:maps_logU_NHI_CIV}, we present the same diagnostic diagram, but using \CIV\ instead of \SiIV. The overall behaviour is very similar. {Finally, we checked that for a given $U$, the models are almost insensitive to $n_{H}$ and $n_{\gamma}$. Only their ratio {matters}.}

\section{Results \label{s:Results}}

{Based on our photo-ionization modeling,}
we locate each component in the $U$–$N(\HI)$ plane using the measured \NV\ and \SiIV\ column densities
{(Fig.~\ref{f:maps_logU_NHI} and \ref{f:maps_logU_NHI_CIV})} and propagate their 1\,$\sigma$ uncertainties accordingly. We assume the metallicity to equate that of the PDLA system, obtained from low-ionization metals. While this is in principle a strong assumption, a shift in metallicity translates into a shift in predicted $N(\HI)$ by the same factor, but has very little effect on the derived ionization parameter.

For weak components where \NV\ is detected but \SiIV\ is not, this approach provides lower limits on $U$. 
When feasible --i.e., for components that are neither saturated nor blended with stronger features-- we then used \CIV\ column density measurements instead of upper-limits on $N(\SiIV)$ to constrain the ionization parameter.

We find that no component occupies the high-$N(\HI)$ region in  {Fig.~\ref{f:maps_logU_NHI}}.
A few components are near the ionization front, but the majority fall within the low $N(\HI)$ regime, well below $\log N(\HI) = 17.5$. In our overall sample, the three bluest \NV\ components towards \Jzzuc\ have their expected \lya\ counterparts shifted outside the DLA trough. For these components, we directly confirm the low associated \HI\ content, as shown in Fig.~\ref{f:fig_J0015_NV_SiIV_SiII}. A Voigt-profile fit, fixing the component's positions, even yields
$\log N(\HI) = 13.7 \pm 0.5 , 13.2 \pm 0.5, 14.2 \pm 0.6$,
which remarkably agrees with the model predictions of 13.8, 13.6, and 14.8, especially given the metallicity dependence of these predicted values and the constraints based on only \NV\ and \CIV. The low \HI\ columns naturally explains the lack of correspondence with low-ionization species, which are expected to arise in more shielded, high-$N(\HI)$ regions.
This behaviour contrasts with the narrow \NV\ components observed in the afterglow spectra of gamma-ray bursts (GRB), which show little to no velocity shift relative to low-ionization metals. In GRB-DLAs, the highly ionized metals are more likely produced in the outer layers of dense clouds situated only a few parsecs from the progenitor star \citep{Prochaska2008}.

 \begin{figure*}
    \centering
    \includegraphics[trim={2.5cm 0 3.65cm 1cm},clip,width=\hsize]{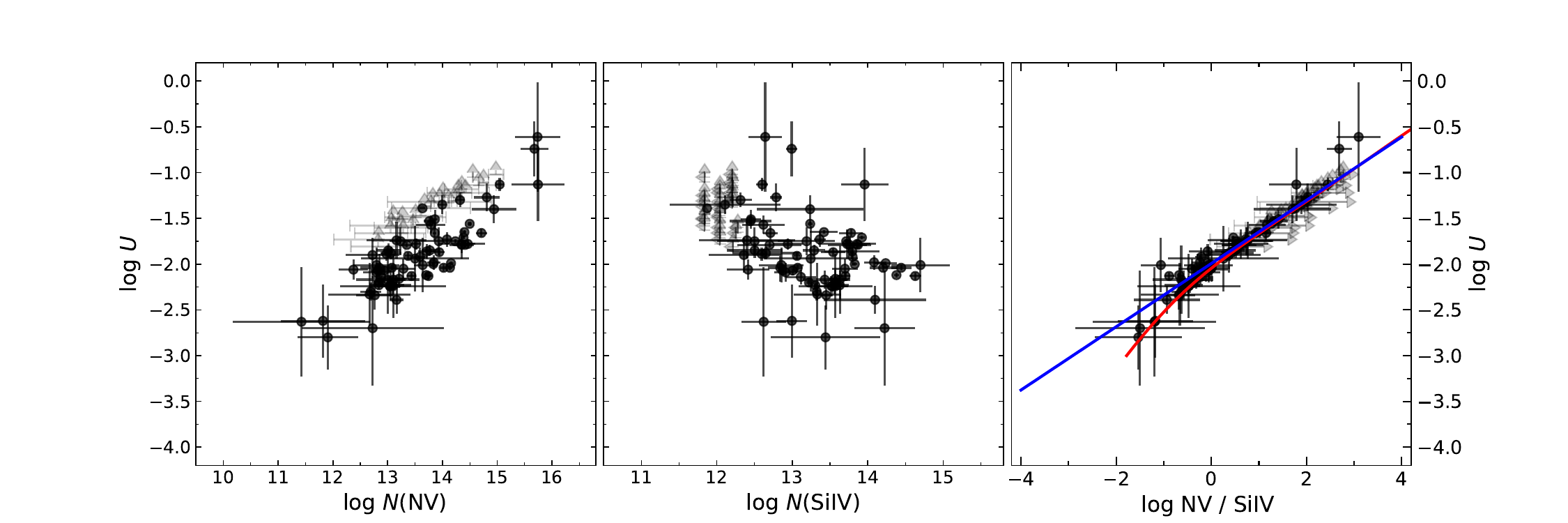} \\
    \caption{
Ionization parameter vs column densities of \NV, \SiIV, and their ratio. Each point represents an individual \NV-bearing component. {Grey arrows represent limits on $U$ for components where \SiIV\ is not detected.}. The red and blue lines in the rightmost panel represent the model-predicted ratio for a fixed $\log N(\HI)=15$ and a simple linear approximation ($\log U \approx 0.4\times\log(\NV/\SiIV) - 2$), respectively. 
}
    \label{f:comparison_NV_SiIV}
\end{figure*}
 
Returning to proximate quasar systems, the \NV\ column density alone would offer an initial indication of the ionization parameter, whereas \SiIV\ does not, see Fig.~\ref{f:comparison_NV_SiIV}.
The \NV/\SiIV\ ratio, however, serves as a more reliable and robust proxy for the ionization parameter. It also eliminates dependencies on metallicity and \HI\ column density for $\log N(\HI) \lesssim 17$.
This relationship is approximately linear in log-space: $\log U \approx 0.4 \times \log(\NV/\SiIV) - 2$, offering a convenient diagnostic tool in the absence of detailed photo-ionization models, provided 
the \HI\ column density remains low. 

\section{\textbf{Discussion}}

\begin{figure*}[]
\centering
\begin{tabular}{cc}
    \includegraphics[height = 0.45\hsize,trim={0.3cm 0 0.25cm 0},clip]{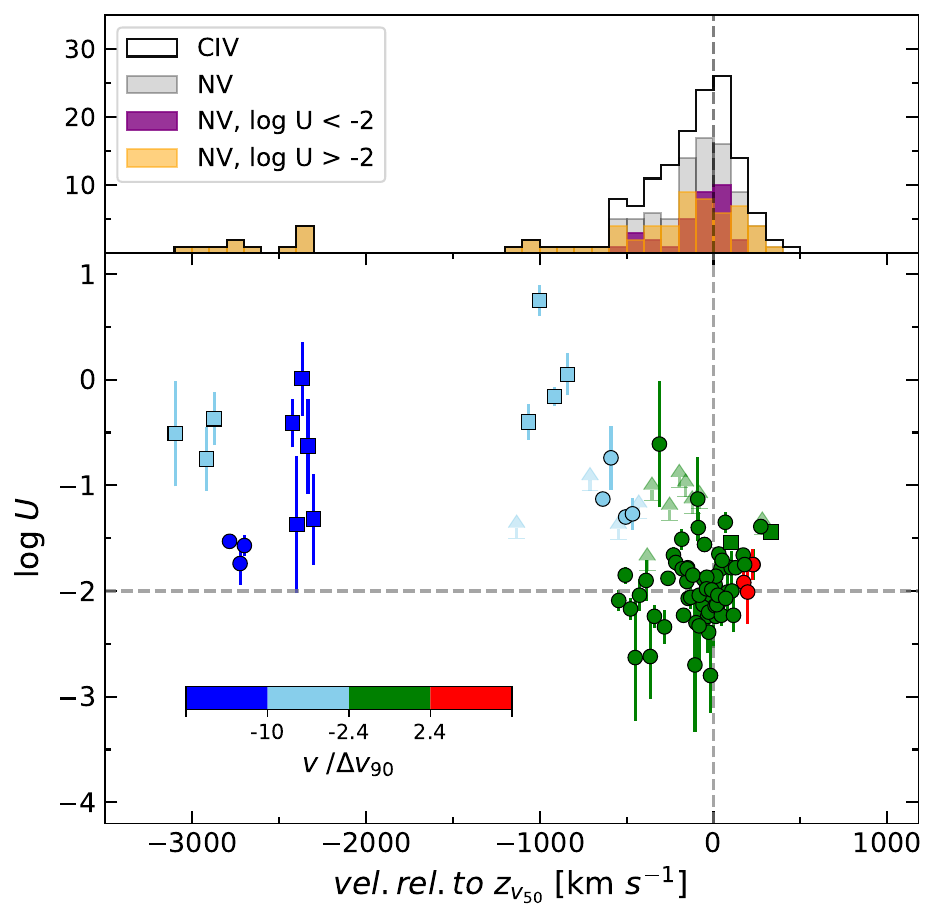} &
    \includegraphics[height = 0.45\hsize,trim={0.8cm 0 0.25cm 0},clip]{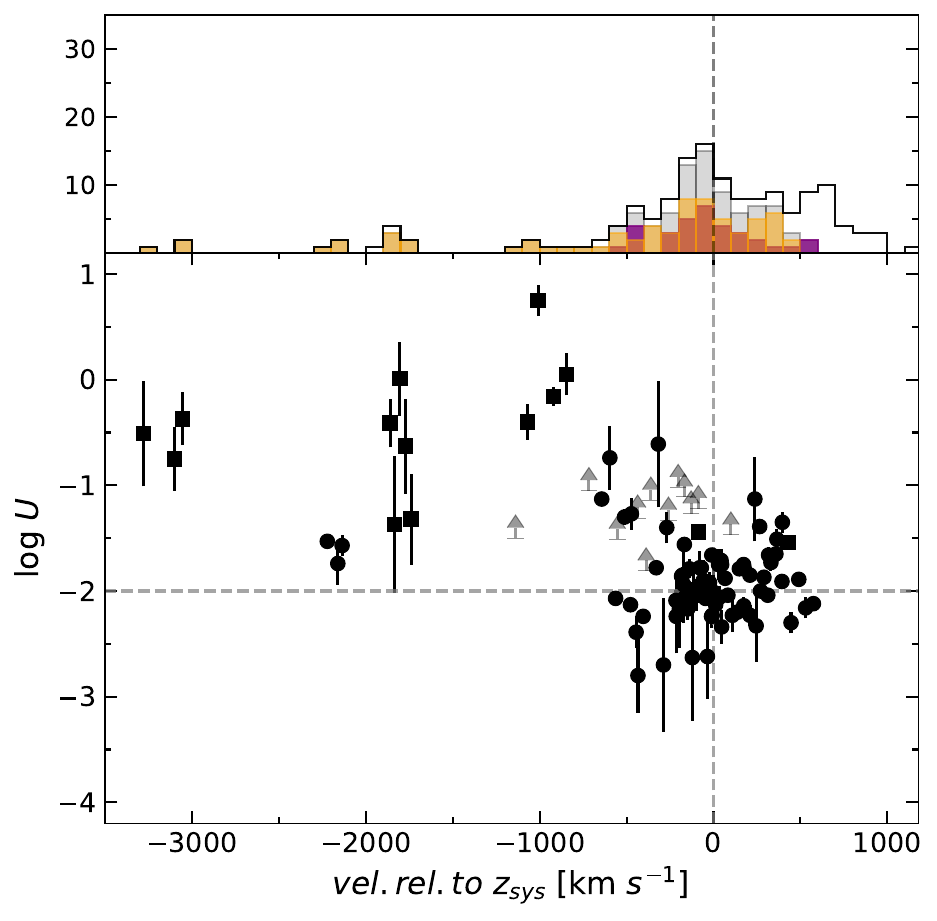}
    \end{tabular}
    \caption{Ionization parameter vs velocity of \NV-bearing components with respect to the bulk of the neutral gas (left panel) or the quasar systemic redshift (right panel). Ionization parameters were derived from \NV\ together with \SiIV\ (circles and lower-limits) or \CIV\  (squares). The top panels show the velocity distribution of \CIV\ (empty histograms) and \NV\ (grey histograms) as well as sub-samples devised upon the derived ionization parameter
    Some \CIV\ components do not exhibit \NV. {In the left panel, points are colour-coded according to their velocity normalised to that of the neutral gas, as used by \citet{Fox2007} to define whether a high-ionization component is likely gravitationally bound ($\abs{v} < 2.4\,\Delta v_{90}$) or not to the DLA (see text).}
    }  
    \label{f:logU_v50_vnorm_vsys} 
\end{figure*}

With ionization {parameter} measurements now available for each component, an important question remains: the location and origin of the observed \NV\ clouds. As previously argued, the \NV\ components are likely located closer to the quasar than the first cloud exhibiting a significant \HI\ column\footnote{
Because of expected clustering around quasars, one might consider the possibility of a second AGN in the field which could 
illuminate the gas
from a different direction than assumed in the models.
We therefore checked that no companion AGN is detected in SDSS in any of the studied field at least out to 200~pkpc at the redshift of the systems.}
Additional insights can be obtained from the ionization stage and relative velocities of the components (see Fig.~\ref{f:logU_v50_vnorm_vsys}).

{In the case of intervening DLAs, \citet{Fox2007} defined an escape velocity $v_{esc}\simeq 2.4 \Delta v_{neutral}$, where $\Delta v_{neutral}$ is the velocity extent measured from absorption lines of singly-ionized metals (i.e. taken here as $\Delta v_{90}$).}
Components exceeding the escape velocity were interpreted as originating from winds.
The left panel of Fig.~\ref{f:logU_v50_vnorm_vsys} shows how the ionization parameter varies across velocity components for the entire sample, relative to the bulk of the neutral gas, and color-coded according to the criterion by \citealt{Fox2007}. 
To define the zero point of the velocity scale, we use $z_{50}$ ({Table~\ref{t:sum}}), which corresponds to the redshift  
where the cumulative optical depth of 
low-ionization metals reaches 50\% of its maximum value. 
This effectively represents the optical depth-weighted centroid of the neutral phase.
We observe that the high-ionization profile extends over a significantly larger velocity range than the low-ionization profile and observe a pronounced asymmetry in the velocity distribution, characterized by a long tail extending toward blue velocities. Coupled with the likelihood that \NV-bearing components are located upstream along the line of sight relative to the neutral gas, this asymmetry suggests that these components are moving away from the quasar, directed toward the bulk of the \HI-absorbing gas.

{This rises also the question of the persistence of \NV\ in potential fast-moving outflows.
The dynamical timescales corresponding to the motion of the absorbing gas can be estimated as $t_{\rm dyn} = d / v$, where $v$ represents the velocity of the \NV\ component ($v < 3000$~\kms) and $d$ is its distance from the AGN.

For the typical quasar luminosity in our sample, the distance can be estimated as $d \sim (10^9~{\rm cm^{-3}} / U / n)^{0.5}$\,pc, resulting in $t_{\rm dyn} > 10^{15} (n / {\rm cm}^{-3})^{-0.5}$\,s for $\log U < -1$, which is relevant to our studies. This timescale is significantly longer than the recombination timescale, $t_{\rm rec} \approx 10^{11} (n / {\rm cm}^{-3})^{-1}$\,s. In other words, the gas reflects the local ionization conditions.

In turn, the ionization state of the gas may or may not follow the AGN flickering or duty cycle, depending on its density \citep[e.g.][]{Rogantini2022}. For example, under different assumptions: (i) for gas with $n= 10^4$\,cm$^{-3}$ (similar to broad absorption line outflows), located at $\sim1$\,kpc (based on the AGN luminosity and $\log U=-1$), the recombination time will be $\lesssim1$\,yr, allowing the gas to follow any variations on timescales longer than this; (ii) for gas with $n=10^{-4}$\,cm$^{-3}$, akin to IGM conditions, the recombination time will be order of $\gtrsim10$\,Myr and the gas will only respond to variations on longer timescales. A more sophisticated time-dependent model would be required to fully capture these variations in the latter case. 

Interestingly, \citet{Oppenheimer2013} emphasized that such fluctuations can significantly impact ions like \NV, which may remain over-ionized for extended periods due to the slow recombination timescales of highly ionized species. This suggests that \NV\ could serve as a tracer of past AGN activity, even after the ionizing source has faded.
However, we lack precise knowledge of the gas number density or the AGN activity history, making it challenging to draw definitive conclusions. That said, the ionization parameters derived from our models likely represent characteristic conditions. 

Additionally, \NV\ can also be produced via collisional ionization in gas with temperatures around $10^{5.3 \pm 0.3}$\,K, \citep[where the ionization fraction peaks under collisional ionization][]{Gnat2007}. These temperatures correspond to thermal $b$-parameters exceeding 15\,\kms, which may be inconsistent with the total width of many of the observed components. Our models also show that radiation from the nearby quasar is sufficient to reproduce the observed features, making it the simplest and preferred scenario.
}

We note that, in spite of the lower ionization potential of \CIV\, that does not require in principle \CIV\ to occur upstream with respect to the \HI\ gas, the corresponding distribution is also skewed towards negative (blue) velocities. 
This suggests that, across our sample, these components are unlikely to originate from winds associated with an external galaxy (i.e., other than the quasar host). If this were 
the case, the velocity distribution would be expected to extend symmetrically also toward positive (red) velocities, as observed in intervening DLAs.
In short, the highly ionized gas seen at significantly blue velocities with respect to the low-ionization metals are likely moving along the line of sight, from the quasar towards the bulk of the \HI\ gas. Interestingly, we can see that the asymmetry in the velocity distribution remains present even when restricting to points within $2.4\times\Delta v_{90}$ (green in the figure),  meaning that the escape velocity criterion by \citet{Fox2007} could also be too restrictive on average.

As we are examining quasar environments, it is interesting to analyze ionization as a function of velocity relative to the quasar's systemic redshift, derived from the quasar emission lines \citepalias[see][]{Noterdaeme2023}, as illustrated in the right panel of Fig.~\ref{f:logU_v50_vnorm_vsys}.
Interestingly, the \CIV\ distribution now extends to redshifted velocities as well. This is primarily attributed to the three most redshifted DLAs (relative to the quasar), where the ionized gas more closely aligns with the neutral components in velocity space. However, the overall distribution of ionized (\CIV) and highly ionized (\NV) gas still shows a significant extent toward negative (blue) velocities. 
Additionally, it is noteworthy that sub-samples defined by their ionization parameter reveal distinct velocity distributions: the highly ionized sample (with $\log U> -2$) includes all components with $v<-750$ $\kms$, while the sample with lower ionization (with $\log U<-2$) shows components centered around zero velocity\footnote{
Part of the observed velocity dispersion could be attributed to uncertainties in the quasar systemic redshift, which is determined with a few hundred \kms\ precision \citepalias{Noterdaeme2023}.}.

We can generally expect that clouds with higher ionization parameters are more likely to be located closer to the quasar than those with lower $U$ values. This is consistent with an origin in quasar outflows, where highly ionized gas exhibits high velocities near the quasar \citep[see also][]{Nestor2008, Balashev2023} and may eventually decelerates to the systemic velocity at greater distances, possibly when reaching the neutral gas, whose illuminated layer could also in principle produce \NV.  

\begin{figure}
    \centering
    \includegraphics[trim={0.2cm 0cm 0.2cm 0cm},clip,width =\hsize]{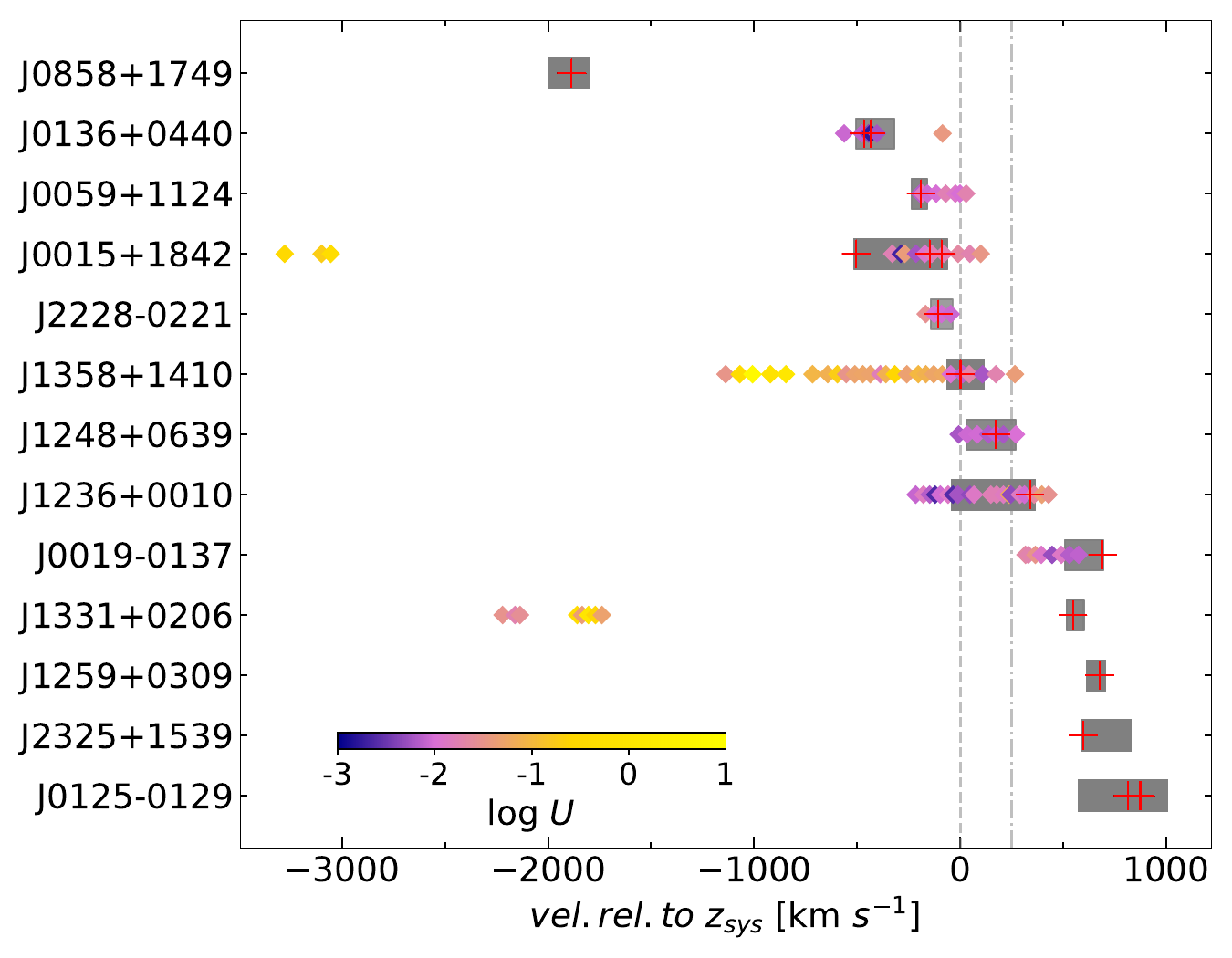}
    \caption{PDLAs in our sample ordered according to their velocity difference between the centroid of the neutral gas (estimated from $z_{50}$) and the systemic redshift of the quasar from optical emission lines. The dashed-dotted lines shows the typical velocity shift of the CO(3-2) emission in 6 systems observed with NOEMA, which may, on average, represent a better measurement of systemic redshift. 
    The grey rectangles depicts the low-ionization $\Delta v_{90}$ range. Red crosses represent the position of H$_2$ components. Diamonds represent \NV\ components, coloured according to the derived ionization parameter.
    }
    \label{f:plot_velocities_emission}
\end{figure}

All these considerations should be taken with caution as we are analyzing the sample as a whole. Fig.~\ref{f:plot_velocities_emission} summarizes the ionization and velocity properties for each quasar individually. Based solely on the \NV\ kinematics, one might interpret \Jzzuc, \Jutch, and \Juttu\ as exhibiting clear signatures of outflowing winds (high blueshifted velocities with high ionization). Indeed, a detailed study of \Jzzuc\ did reveal the presence of a multi-phase, multi-scale outflow in that system \citep{Noterdaeme2021}.

The PDLA towards \Jzhch\ is strongly blueshifted but lacks highly ionized gas. It is therefore plausible that this PDLA is an intervening system along the line of sight, and the $-2000$~\kms\ offset may originate from the Hubble flow. The presence of one such intervening system in our sample aligns with the observed excess of $\sim 10-20$ times more H$_2$ systems close to the quasar redshift compared to what is expected from intervening statistics \citepalias{Noterdaeme2019}. This suggests that the velocity range used to associate a DLA with the quasar could be restricted to about $\sim 1000~\kms$ from the systemic redshift, while highly ionized gas can exhibit much larger velocities \citep[see also][for similar conclusion from \CIV\ and \MgII\ statistics]{Chen2017}.

Conversely, the three PDLAs with the most redshifted components show no evidence of \NV. A comprehensive study involving X-shooter, Subaru, and ALMA data of {one of them} demonstrated that the absorption arises from a merging companion galaxy \citep{Balashev2024}.

Since molecular hydrogen and highly ionized nitrogen probe different phases and velocities relative to the quasar, one may wonder whether the presence of \NV\ is related to the selection based on proximate \HH\ absorption. Based on a search among regular quasars, \citet{Perrotta2016} reached the same conclusion that \NV\ are closely associated with the quasar 
and likely originating from outflowing gas. Interestingly, their study, also based on X-shooter data, revealed the presence of \NV\ absorption in about a third of quasars (although about half of \CIV\ absorbers within $\pm$1000\,\kms\ in their sample have also \NV), when here the fraction is about twice higher. 
This result suggests that, although surprising, the selection based on \HH\ absorption appears to enhance the likelihood of detecting \NV\ compared to the general quasar population.
This excess in the \NV\ detection rate is approximately 70$\%$ in \HH\ - bearing PDLAs, significantly higher than the 13$\%$ observed in intervening DLAs \citep{Fox2009} and the roughly 30$\%$ found in regular quasars \citep{Perrotta2016}.

This may be an indirect effect: proximate \HH\ systems could select quasars surrounded by significant amounts of neutral and molecular gas, possibly witnessing the formation of these objects, as these are expected to be triggered by galaxy interactions {and to feature strong outflows}. In fact, the three \HH\ systems in our sample that have been studied in more details using ancillary data present remarkable properties: one is a possible post-merger system with a large reservoir of molecular gas with high velocity dispersion \citep{Noterdaeme2021b}, another lives in a rich galactic environment as traced by \lya\ emission \citep{Urbina2024}, and the third is part of a major merging system \citep{Balashev2024}.
Additionally, six out of the 13 systems have been observed with NOEMA, resulting in six detections of CO(3-2) emission \citepalias{Noterdaeme2023}.

Interestingly, the most redshifted PDLAs (likely tracing \HI\ and \HH\ gas infalling toward the quasars) show no sign of outflow as indicated by the absence of \NV\ absorption. This contrasts with the more blueshifted systems or those around the quasar systemic velocity, which do display clear outflow signatures. This
suggests that while the highly ionized phase (\NV) is decoupled from the \HI\ phase, outflowing gas is primarily detected when PDLAs have velocities either moving towards the observer (blueshifted) or close to the quasar systemic velocity.

\section{Conclusion\label{s:conclusion}}

In this work, we investigated the presence of four times ionized nitrogen (\NV) in a sample of 13 quasars selected based on the detection of \HH\ absorption.
A crucial aspect of our study was the use of medium-high resolution spectroscopy, which enabled us to examine the properties of \NV\ across the velocity profile. We confirmed  previous claims that \NV\ is an effective tracer of quasar proximity \citep[e.g.][]{Ganguly2013,Perrotta2018} and quantify this by deriving the ionization parameter. Notably, we find higher ionization parameters in components at large blue velocity shifts from the quasar.
We also presented a straightforward and practical method for estimating this {ionization} parameter using the \NV-to-\SiIV\ ratio.

We conjecture that proximate \HH\ absorbers trace quasars during a key stage of their evolution, capturing the phase when galaxy interactions drive significant amounts of molecular gas toward the central engine, fueling it, while feedback in the form of outflows begins to expel gas outward.

Whether the different properties observed in our sample are primarily due to evolutionary processes—such as a merger phase followed by outflows—or due to orientation, with infalling gas and outflows occurring in distinct directions, remains an open question. This underlines the need for more systematic and detailed studies of quasars with proximate \HH\ absorbers, examining not only the physical, chemical, and kinematical properties of the gas but also the emission characteristics of the quasars themselves and their surrounding galactic environments.

\begin{acknowledgement}
RC and PN gratefully acknowledge support from the French Chilean Laboratory for Astronomy. RC and SB are thankful to IAP for hospitality during part of the time this work was done, and so is PN with the Astronomy Departments of Universidad de Chile. SB is supported by RSF grant 23-12-00166. RC and SL acknowledges support by FONDECYT grant 1231187. The research leading to these results received support from the {\sl Agence Nationale de la Recherche}, Project ANR-17-CE31-0011 "HIH2" (PI. Noterdaeme). 
\end{acknowledgement}

\bibliographystyle{aa}

\begin{appendix}
\section{Fitting results 
\label{s:Appendix}}
\label{s:indv_spectra}

\begin{figure*}[h]
    \includegraphics[trim={1.2cm 1.2cm 0 0},clip,width =  \hsize]{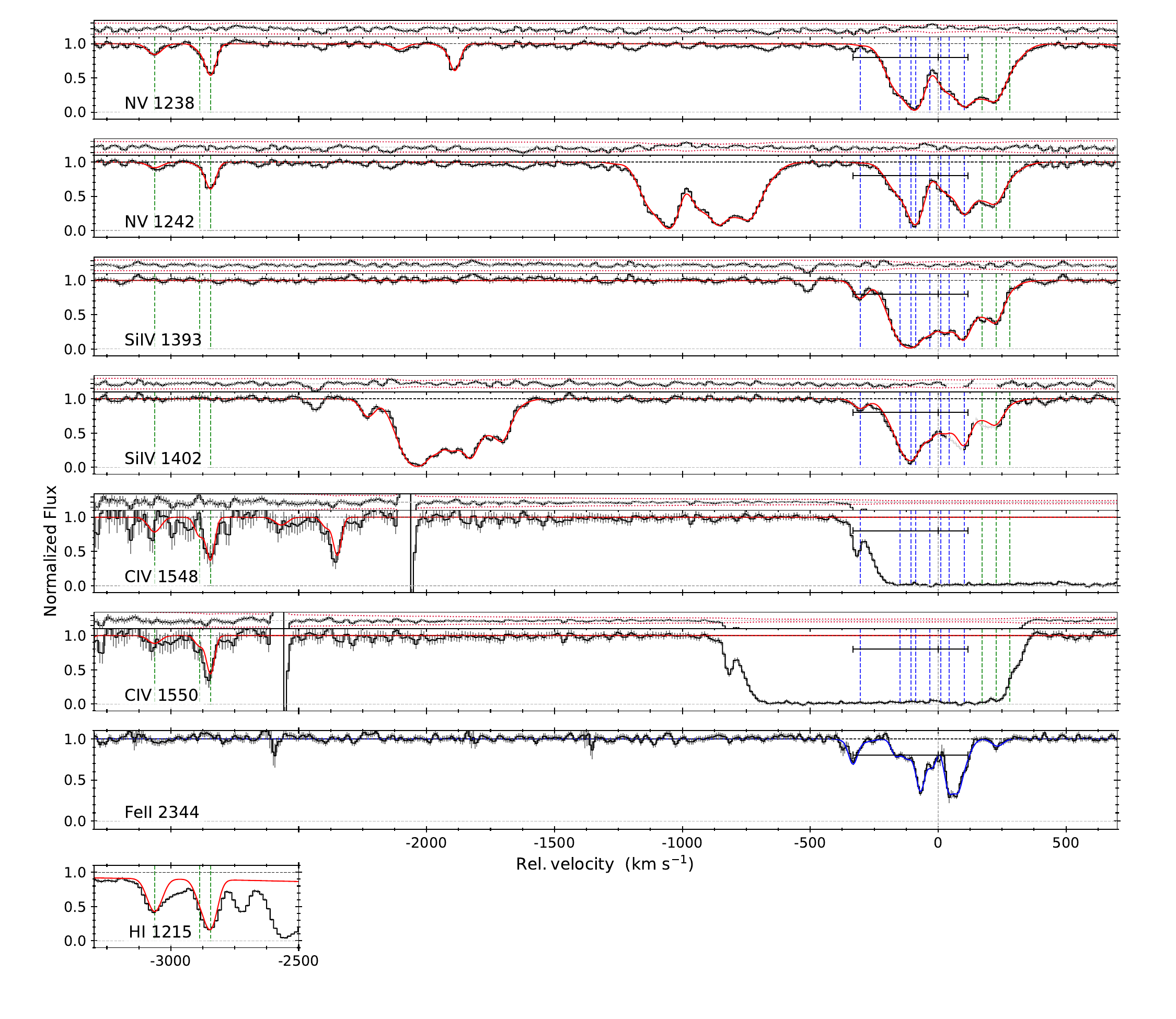}
    \caption{\NV, \SiIV\ and \CIV\ absorption lines in the PDLA towards \Jzzuc. The zero of the velocity scale is set to $z_{50}$ = 2.628810.
    The best Voigt-profile model is shown in red. For comparison, we also show the profile of a low-ionization metal line, along with its best-fit model (blue) from \citet{Noterdaeme2021}; \citetalias{Noterdaeme2023} or \cite{Balashev2019}, as appropriate. The black horizontal segment depicts the $\Delta v_{90}$ range of the neutral gas, with tick marks at 5\%, 50\% and 95\% ($v_5$, $v_{50}$ and $v_{95}$) of the cumulative optical depth.
    The blue vertical line represents high-ionization metal components within $\Delta v_{90}$ while green vertical lines mark those outside this range. The short bottom panel shows the region where individual \HI\ \lya\ components can be clearly identified, i.e. not blended within the saturated DLA profile.}      
    \label{f:fig_J0015_NV_SiIV_SiII}
\end{figure*}

\begin{figure}[h]
    \includegraphics[width = 3.4in]{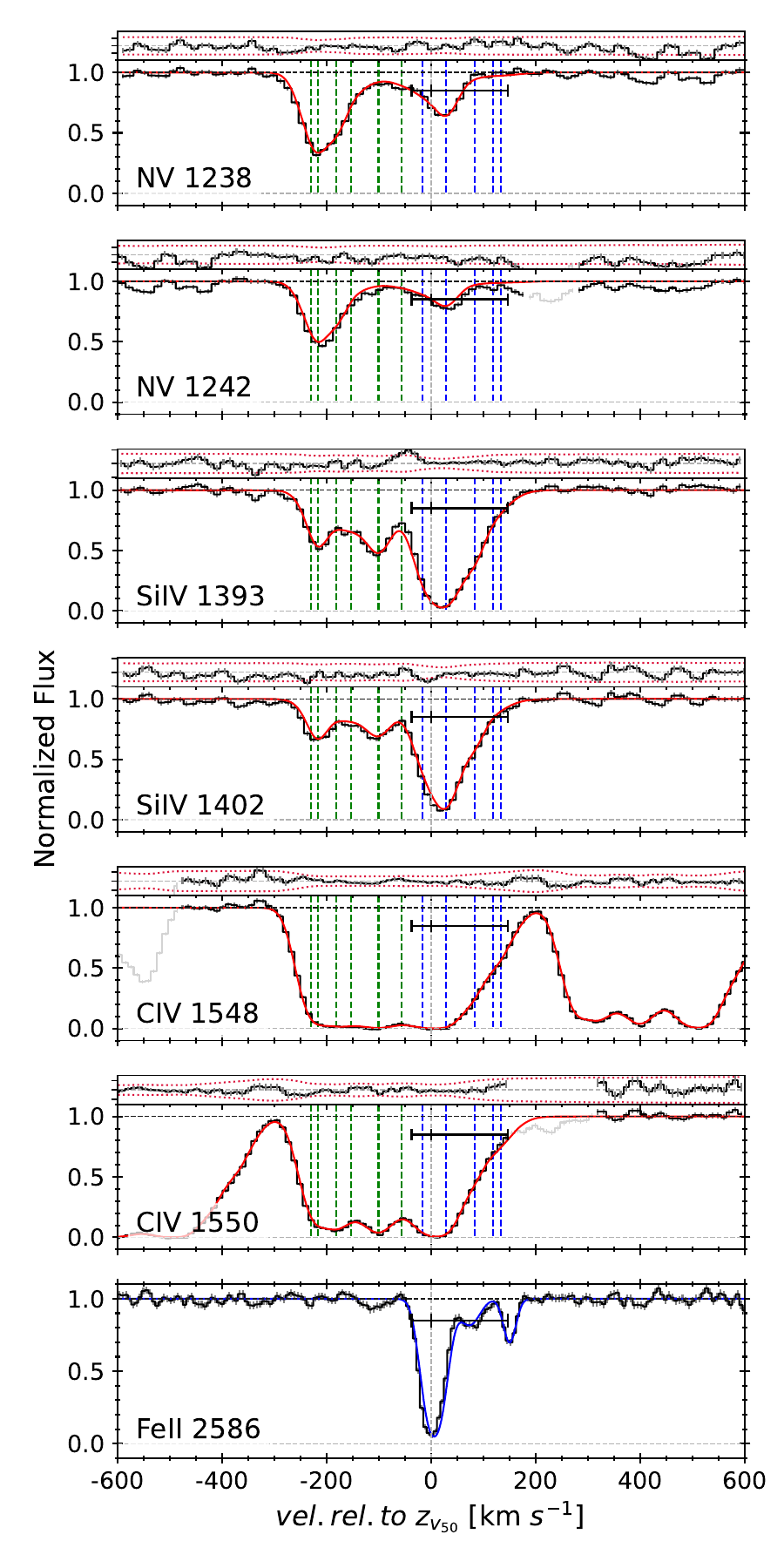}
    \caption{Same as Fig. \ref{f:fig_J0015_NV_SiIV_SiII} for \Jzzun. $z_{50}$ = 2.528134.}      
    \label{f:fig_J0019_NV_SiIV_CIV}
\end{figure}

\begin{figure}[h]
\includegraphics[width = 3.4in]{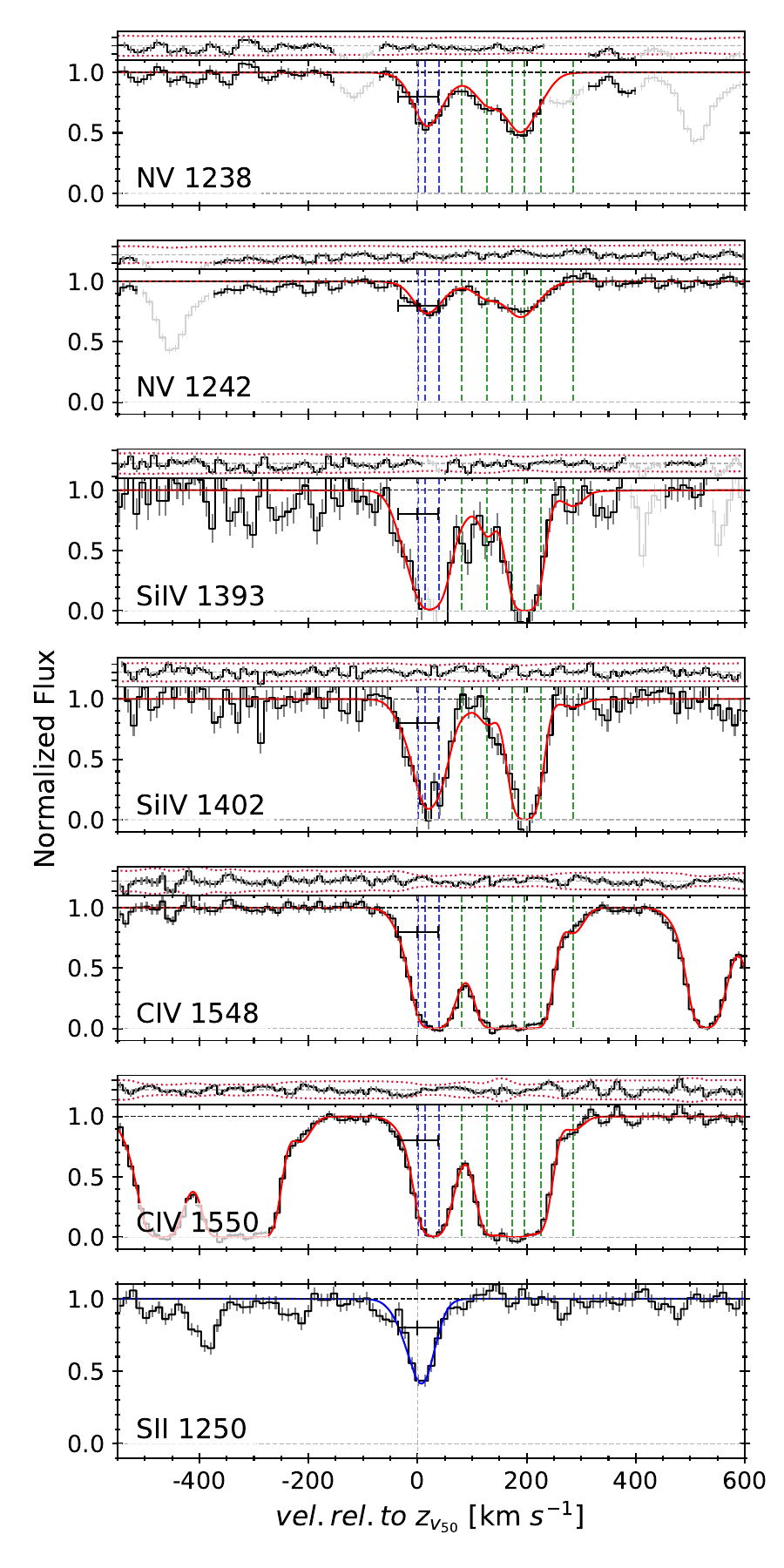}
    \caption{Same as Fig. \ref{f:fig_J0015_NV_SiIV_SiII} for \Jzzcn. $z_{50}$ = 3.034242.}
    \label{f:fig_J0059_NV_SiIV_CIV}
\end{figure}

\begin{figure}[h]
    \includegraphics[width = 3.4in]{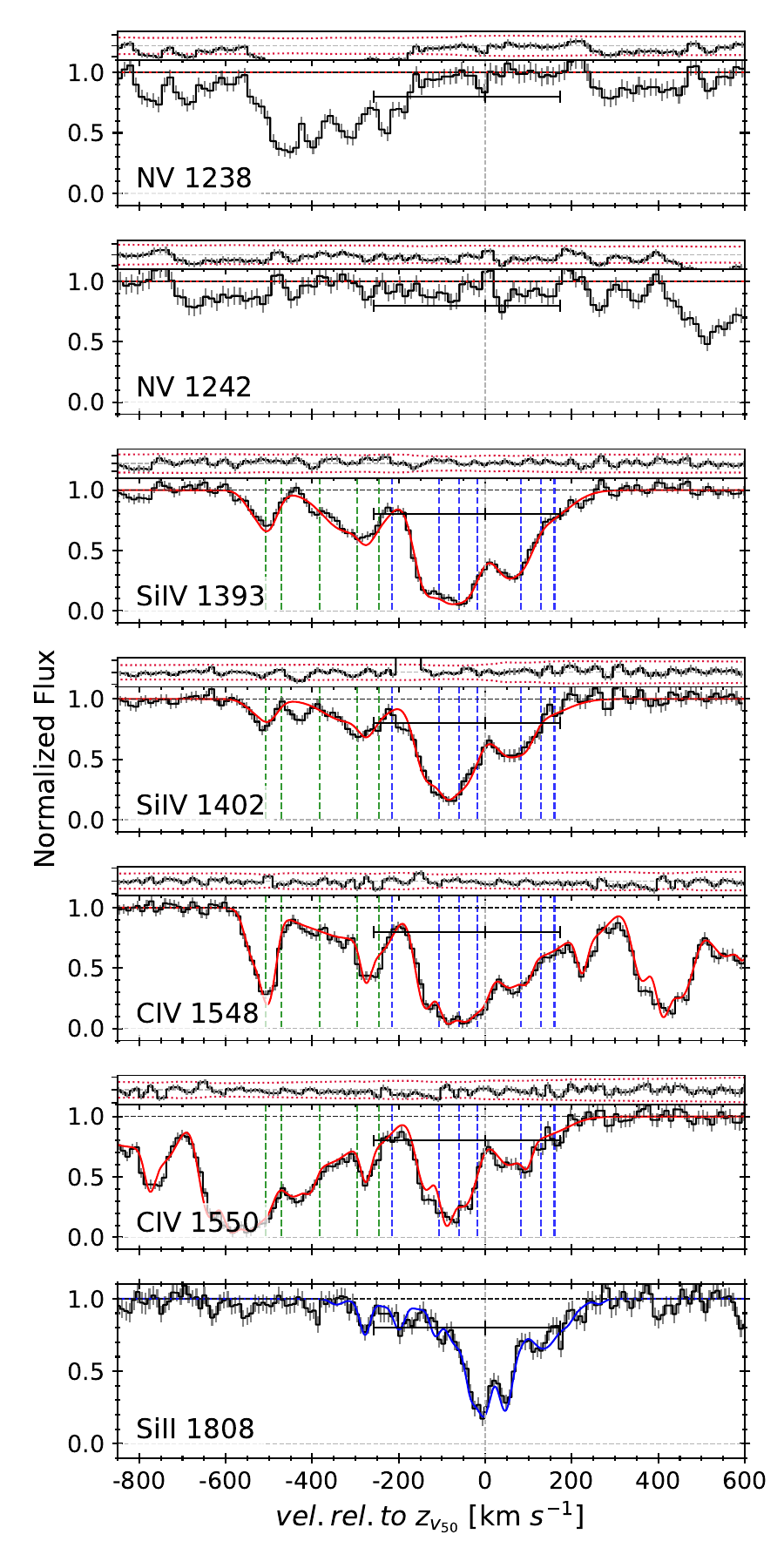} 
    \caption{Same as Fig. \ref{f:fig_J0015_NV_SiIV_SiII} for \Jzudc. $z_{50}$ = 2.663585.}      
    \label{f:fig_J0125_NV_SiIV_CIV}
\end{figure}

\begin{figure}[h]
    \includegraphics[width = 3.4in]{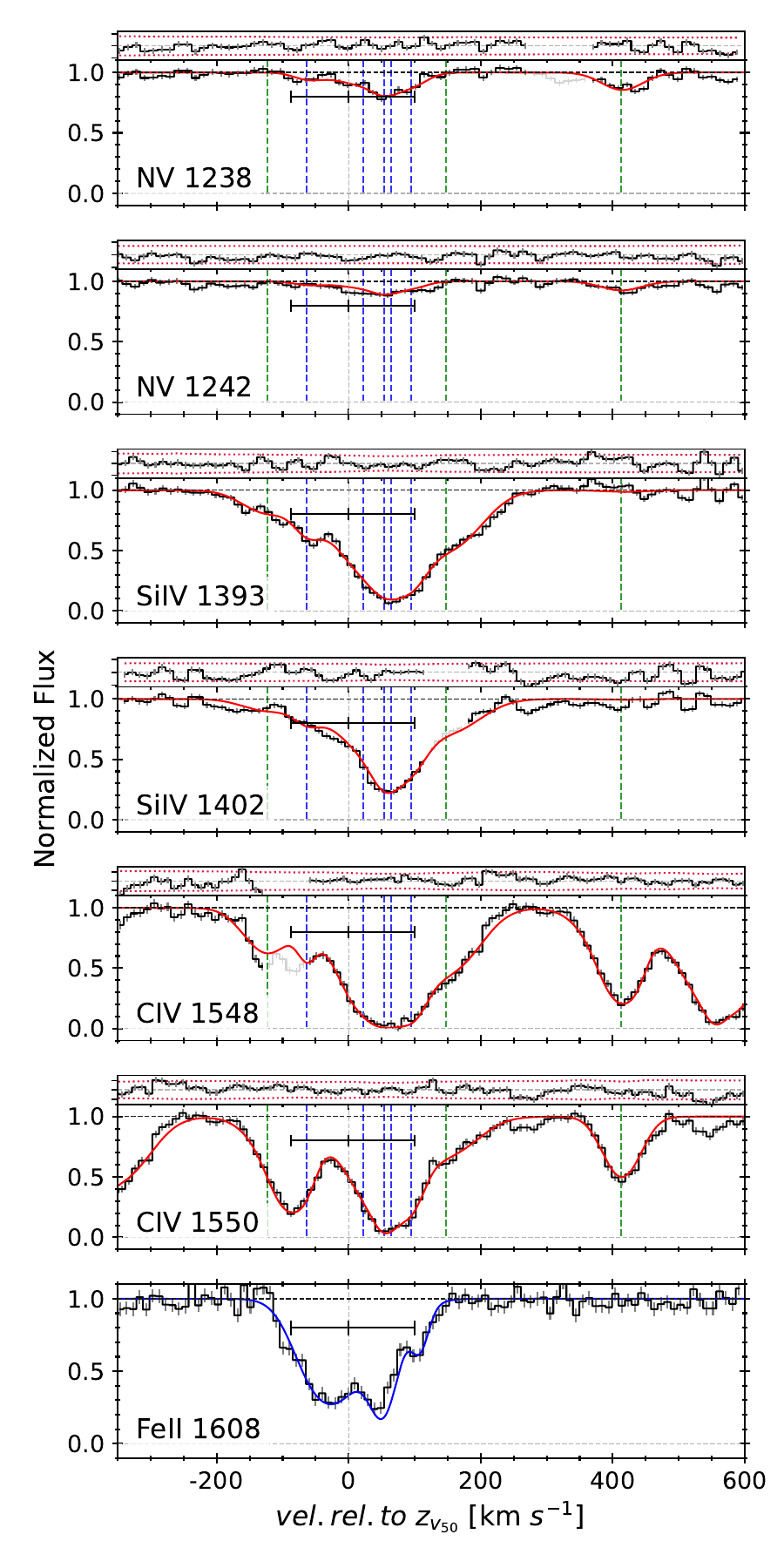}
    \caption{Same as Fig. \ref{f:fig_J0015_NV_SiIV_SiII} for \Jzuts. $z_{50}$ = 2.77861.}
    \label{f:fig_J0136_NV_SiIV_CIV}
\end{figure}

\begin{figure}[h]
    \includegraphics[width = 3.4in]{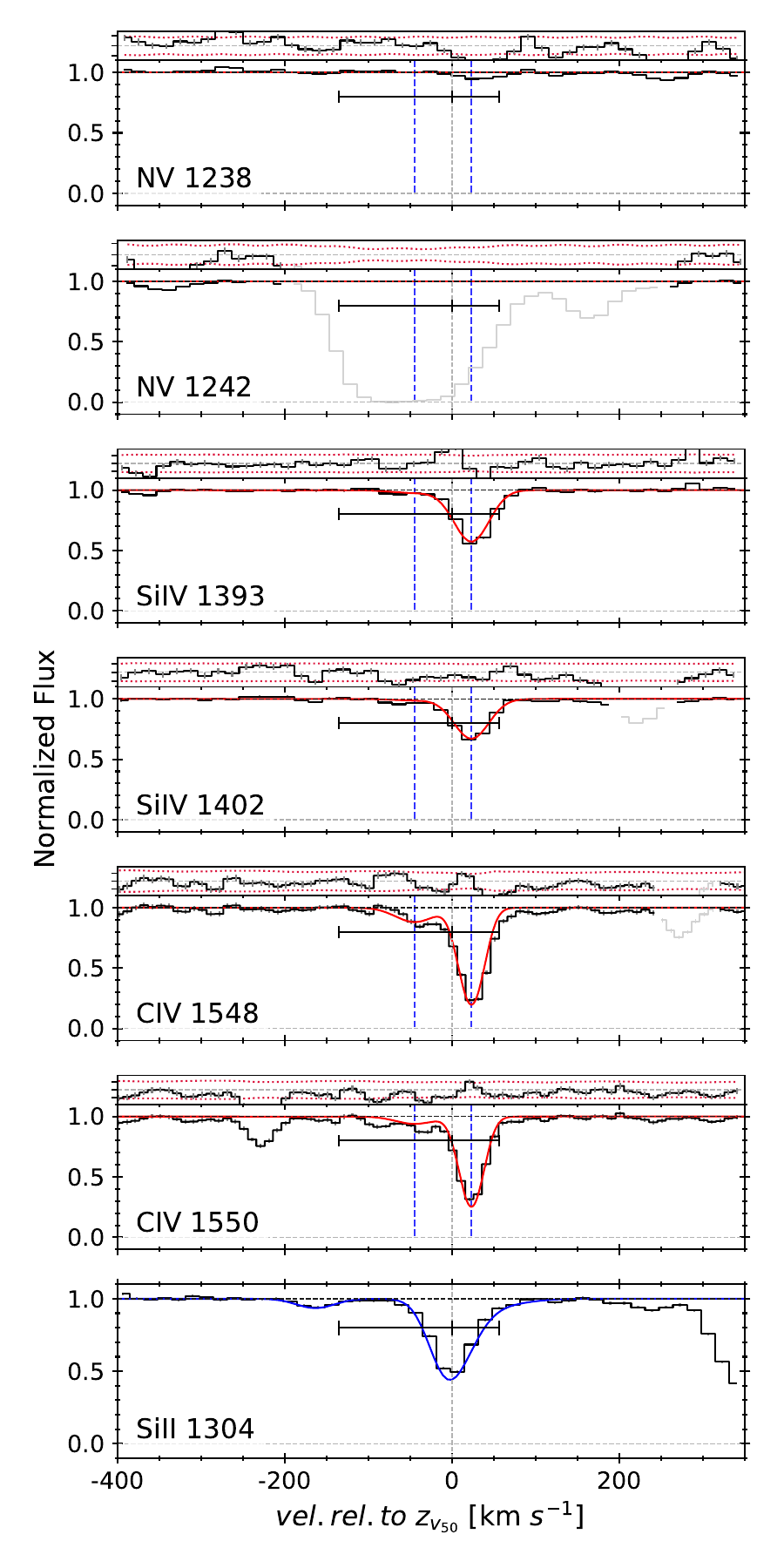} 
    \caption{Same as Fig. \ref{f:fig_J0015_NV_SiIV_SiII} for \Jzhch. $z_{50}$ = 2.625366.   
    \label{f:fig_J0858_NV_SiIV_CIV}}
\end{figure}

\begin{figure}[h]
    \includegraphics[width = 3.4in]{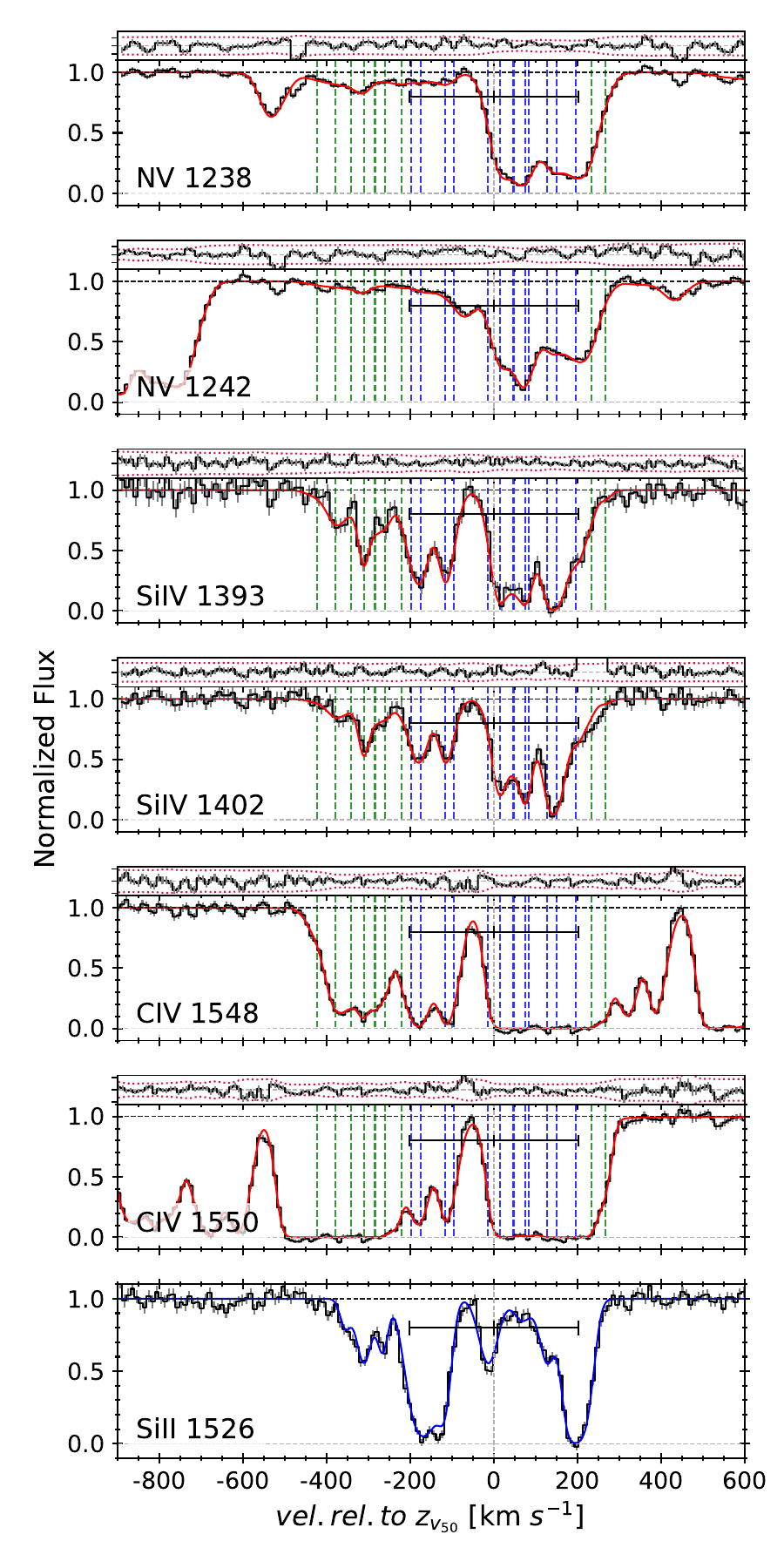}
    \caption{Same as Fig. \ref{f:fig_J0015_NV_SiIV_SiII} for \Judts. $z_{v50}$ =  3.030589}  
    \label{f:fig_J1236_NV_SiIV_CIV}
\end{figure}

\begin{figure}[h]
    \includegraphics[width = 3.4in]{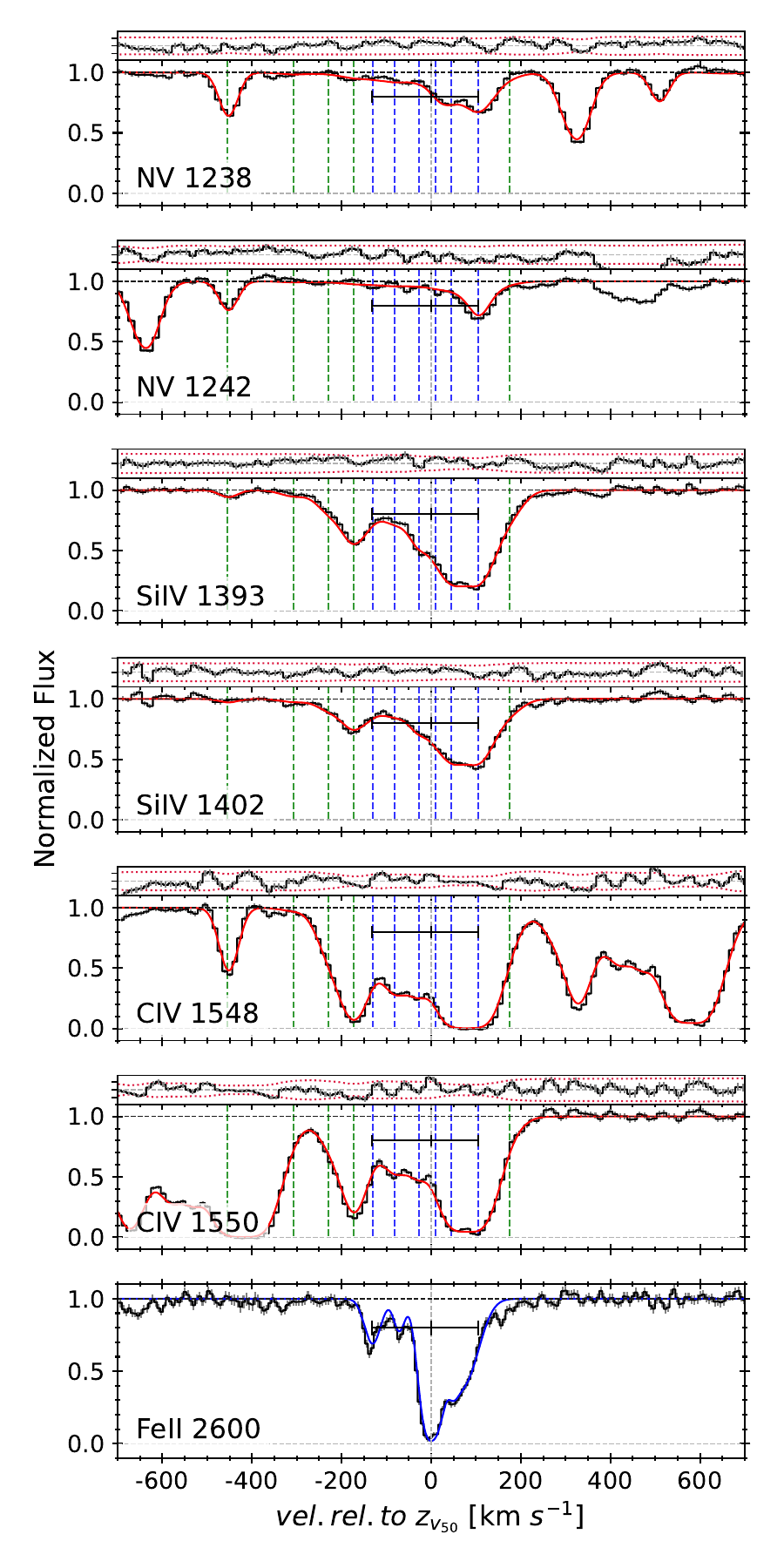} 
    \caption{Same as Fig. \ref{f:fig_J0015_NV_SiIV_SiII} for \Judqh. $z_{50}$ = 2.529240.}
    \label{f:fig_J1248_NV_SiIV_CIV}
\end{figure}

\begin{figure}[h]
    \includegraphics[width = 3.4in]{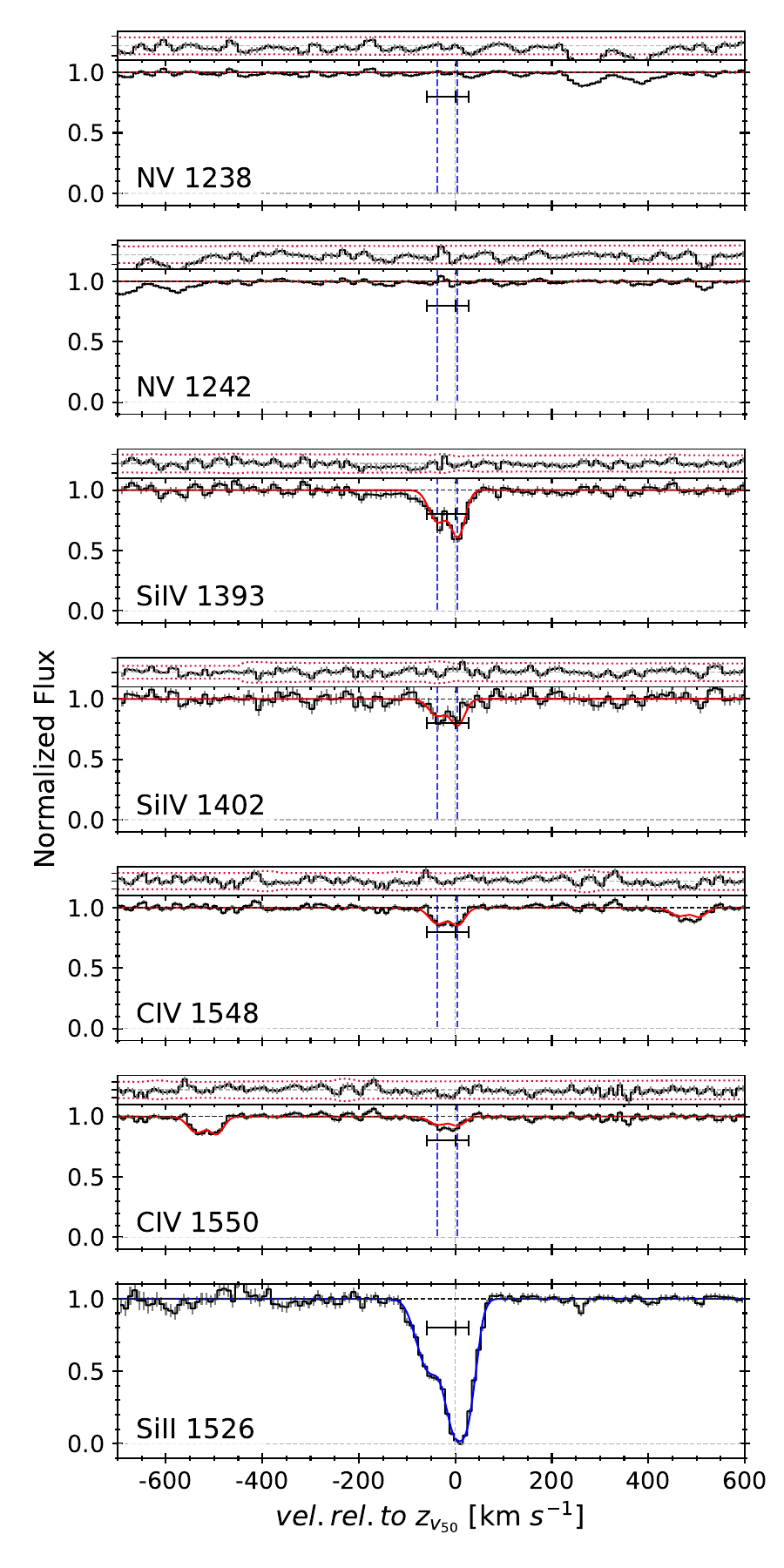}
    \caption{Same as Fig. \ref{f:fig_J0015_NV_SiIV_SiII} for \Judcn. $z_{50}$ = 3.246062.}      
    \label{f:fig_J1259_NV_SiIV_CIV}
\end{figure}

\begin{figure*}[h]
    \centering
    \includegraphics[trim={1.2cm 0.2cm 0 0},clip,width=\hsize]{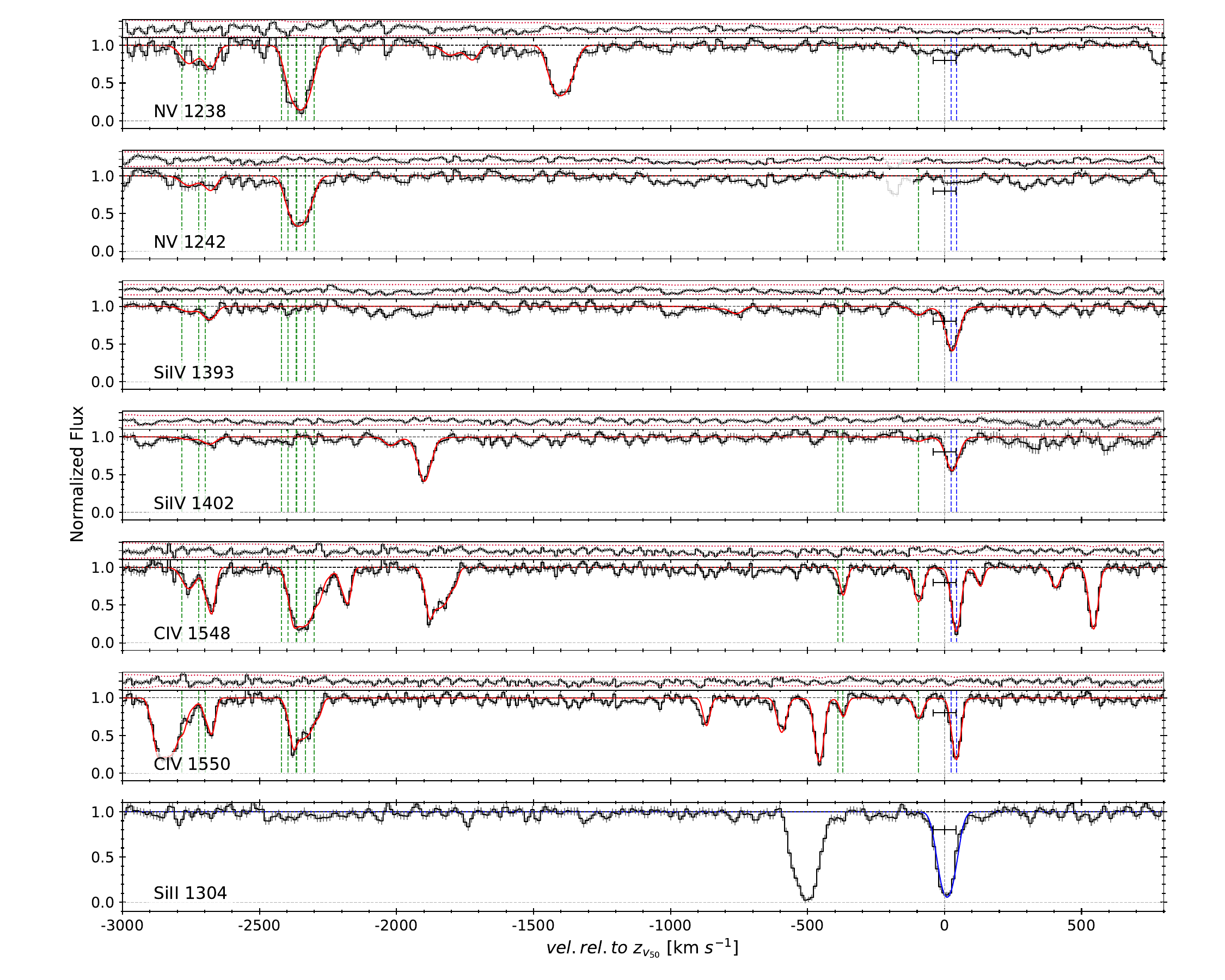} 
    \caption{Same as Fig. \ref{f:fig_J0015_NV_SiIV_SiII} for \Juttu. $z_{50}$ = 2.92186.}      
    \label{f:fig_J1331_NV_SiIV_CIV}
\end{figure*}

\begin{figure}[h]
    \includegraphics[width = 3.4in]{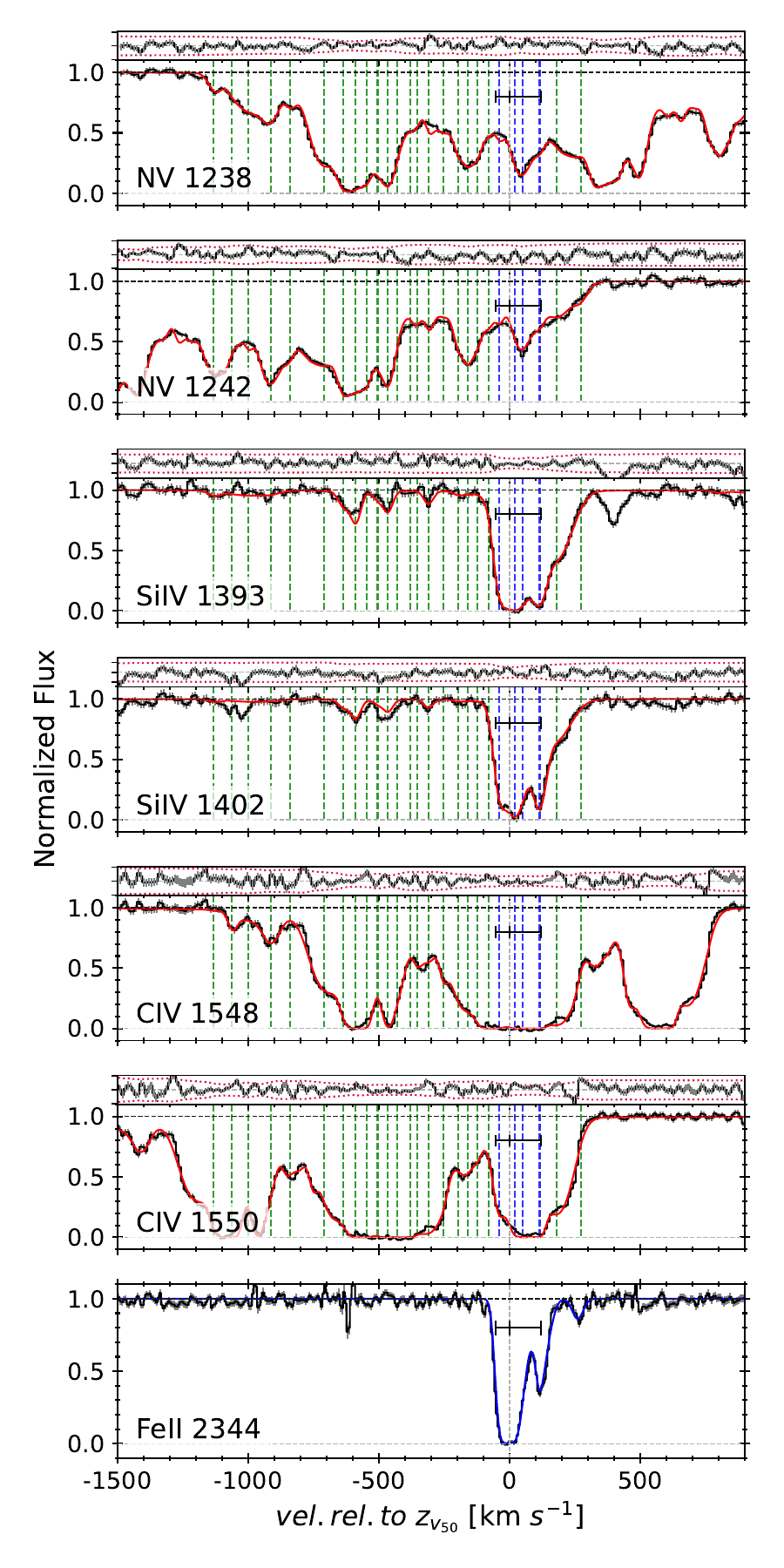}
    \caption{Same as Fig. \ref{f:fig_J0015_NV_SiIV_SiII} for \Jutch. $z_{50}$ = 2.892517.}      
    \label{f:fig_J1358_NV_SiIV_CIV}
\end{figure}

\begin{figure}[h]
    \includegraphics[width = 3.4in]{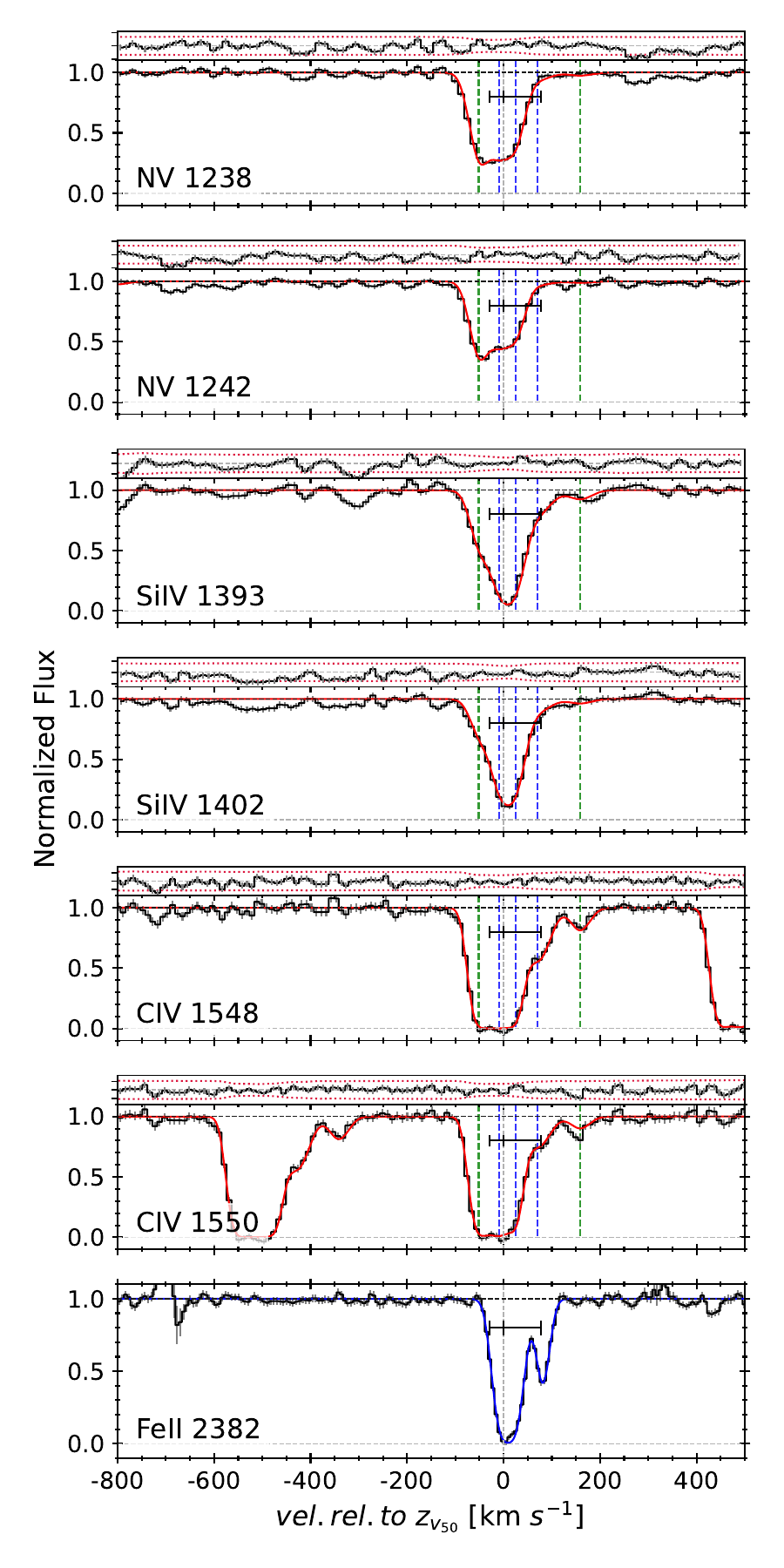}
    \caption{Same as Fig. \ref{f:fig_J0015_NV_SiIV_SiII} for \Jdddh. $z_{50}$ = 2.769043.}      
    \label{f:fig_J2228_NV_SiIV_CIV}
\end{figure}

\begin{figure}[h]
    \includegraphics[width = 3.4in]{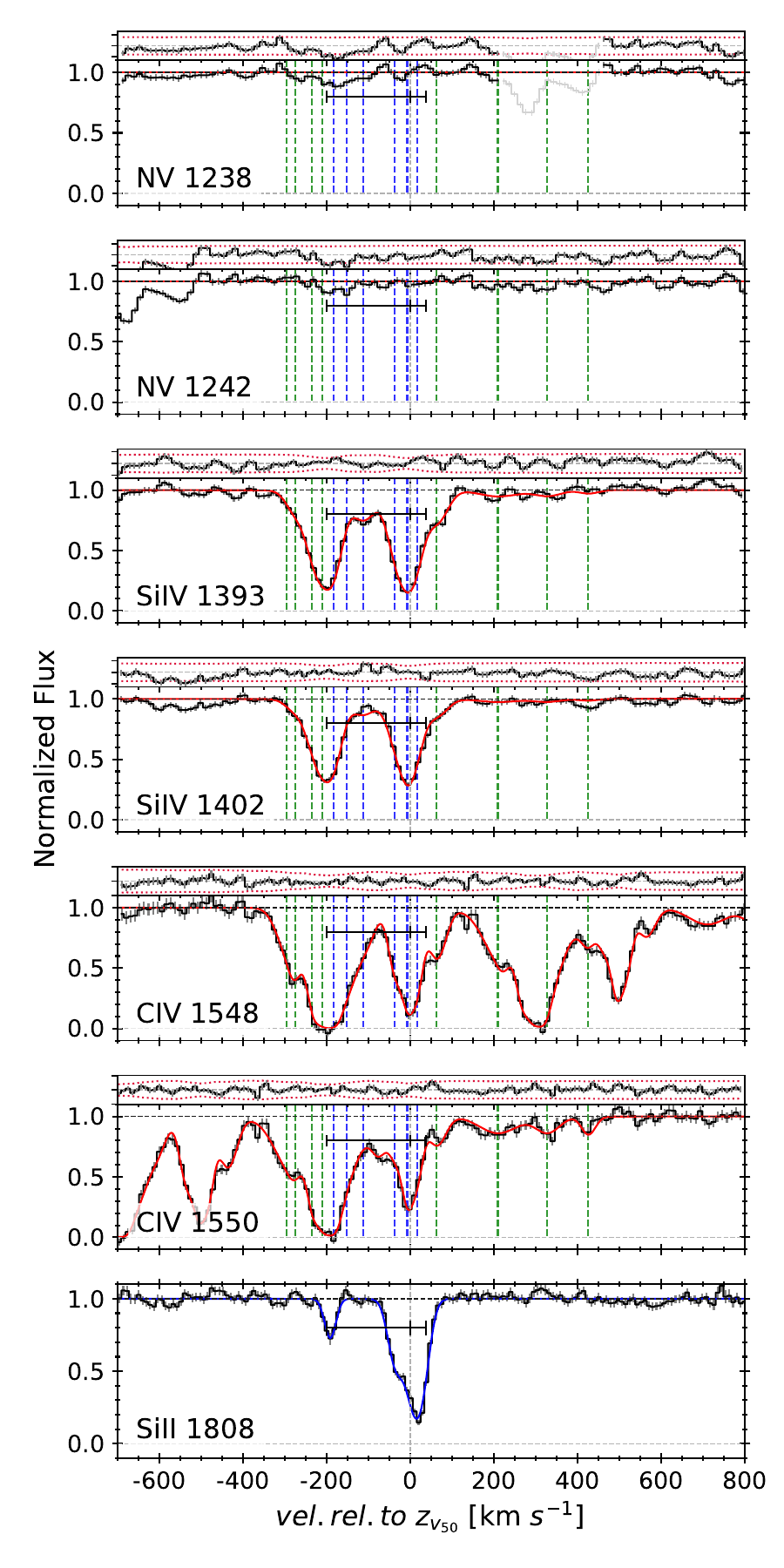}
    \caption{Same as Fig. \ref{f:fig_J0015_NV_SiIV_SiII} for \Jdtdc. $z_{50}$ = 2.616719.}      
    \label{f:fig_J2325_NV_SiIV_CIV}
\end{figure}

\clearpage

\onecolumn
\begin{longtable}{ccc ccc|cc}
\caption{Column densities and ionization parameters in \NV-bearing components.}\\ 
   \hline  \hline 
   Quasar & $z$ & $b$ & log $N(\NV)$ & log $N(\SiIV)$ & log $N(\CIV)$ & \multicolumn{2}{c}{$\log U$} \\
          &     & (\kms)     & [\cmsq] & [\cmsq] & [\cmsq] &  from \NV\ and \SiIV &  from \NV\ and \CIV \\ 

  \hline 
  \endfirsthead
   \caption{continued.}\\
  \hline \hline
     Quasar & $z$ & $b$ & log N(\NV) & log N(\SiIV) & log N(\CIV) & \multicolumn{2}{c}{$\log U$} \\
          &     & [\kms]     & [\cmsq] & [\cmsq] & [\cmsq] &  from \NV\ and \SiIV &  from \NV\ and \CIV \\ 
 
  \hline 

  \endhead
  
  \hline
  \endfoot

  \hline 
  \endlastfoot

 \Jzzuc
& 2.591726          &   25  $\pm$    21 &   13.24  $\pm$  0.49 &  <12.04             &  13.07  $\pm$  1.05   &   > -1.60             &     -0.51   $\pm$  0.50  \\
& 2.593851          &    8  $\pm$     3 &   13.05  $\pm$  0.65 &  <12.04             &  13.03  $\pm$  0.66   &   > -1.66             &     -0.75   $\pm$  0.30  \\     
& 2.594378          &   10  $\pm$     5 &   14.29  $\pm$  0.57 &  <12.04             &  14.04  $\pm$  0.31   &   > -1.22             &     -0.37   $\pm$  0.25  \\     

& 2.627014$\dagger$ &   58  $\pm$    12 &   14.46  $\pm$  0.13 &  13.85  $\pm$  0.18 &  ---                  &    -1.78  $\pm$  0.05 &     ---                  \\     
& 2.627520$\dagger$ &   40  $\pm$    21 &   12.72  $\pm$  1.30 &  14.22  $\pm$  0.40 &  ---                  &    -2.70  $\pm$  0.63 &     ---                  \\     
& 2.627750$\dagger$ &   22  $\pm$    18 &   14.94  $\pm$  0.41 &  13.23  $\pm$  0.70 &  ---                  &    -1.40  $\pm$  0.15 &     ---                  \\     
& 2.628406$\dagger$ &    90  $\pm$   19 &   13.10  $\pm$  0.97 &  13.57  $\pm$  0.48 &  ---                  &    -2.24  $\pm$  0.35 &     ---                  \\     
& 2.628926$\dagger$ &   29  $\pm$    29 &   13.50  $\pm$  0.98 &  13.24  $\pm$  0.62 &  ---                  &    -1.94  $\pm$  0.36 &     ---                  \\     
& 2.629328$\dagger$ &   56  $\pm$    54 &   14.35  $\pm$  0.22 &  13.75  $\pm$  0.28 &  ---                  &    -1.79  $\pm$  0.09 &     ---                  \\     
& 2.630040$\dagger$ &   16  $\pm$    24 &   14.35  $\pm$  0.43 &  13.72  $\pm$  0.38 &  ---                  &    -1.78  $\pm$  0.16 &     ---                  \\     
& 2.630881          &   87  $\pm$    30 &   14.71  $\pm$  0.10 &  13.77  $\pm$  0.15 &  ---                  &    -1.66  $\pm$  0.04 &     ---                  \\     
& 2.631568          &   21  $\pm$    19 &   13.93  $\pm$  0.41 &  13.18  $\pm$  0.38 &  ---                  &    -1.75  $\pm$  0.15 &     ---                  \\     
& 2.632211          &   55  $\pm$    54 &   13.60  $\pm$  0.39 &  <12.04             &  ---                  &    >-1.47             &     ---                  \\

 \hline
\Jzzun 
& 2.525427           &    22 $\pm$    1  &  13.86  $\pm$  0.03  &   12.71  $\pm$  0.09  &   14.35  $\pm$  0.02   &   -1.60  $\pm$   0.08     &       -1.35   $\pm$   0.05   \\
& 2.525585           &     7 $\pm$    1  &  14.08  $\pm$  0.09  &   13.36  $\pm$  0.07  &   14.91  $\pm$  0.30   &   -1.73  $\pm$   0.08     &       -1.62   $\pm$   0.22   \\
& 2.525989           &     8 $\pm$    2  &  13.86  $\pm$  0.04  &   12.45  $\pm$  0.13  &   15.58  $\pm$  0.25   &   -1.51  $\pm$   0.10     &       -2.15   $\pm$   0.15   \\
& 2.526333           &    28 $\pm$   12  &  13.37  $\pm$  0.05  &   13.06  $\pm$  0.03  &   14.37  $\pm$  0.02   &   -1.91  $\pm$   0.04     &       -1.80   $\pm$   0.08   \\
& 2.526948           &    21 $\pm$    5  &  12.69  $\pm$  0.19  &   13.32  $\pm$  0.02  &   14.81  $\pm$  0.06   &   -2.30  $\pm$   0.10     &       -2.55   $\pm$   0.45   \\
& 2.527473           &    20 $\pm$   22  &  12.97  $\pm$  0.10  &   12.59  $\pm$  0.07  &   14.14  $\pm$  0.05   &   -1.89  $\pm$   0.07     &       -1.95   $\pm$   0.15   \\
& 2.527938$\dagger$  &    17 $\pm$    5  &  13.21  $\pm$  0.14  &   13.57  $\pm$  0.08  &   14.65  $\pm$  0.15   &   -2.16  $\pm$   0.10     &       -2.18   $\pm$   0.13   \\
& 2.528463$\dagger$  &    20 $\pm$    2  &  13.70  $\pm$  0.02  &   14.38  $\pm$  0.03  &   15.20  $\pm$  0.05   &   -2.12  $\pm$   0.03     &       -2.09   $\pm$   0.06   \\

\hline
\Jzzcn 
& 3.034257$\dagger$  &    35 $\pm$   9  &  12.99  $\pm$  0.581  &   13.63  $\pm$  0.14  &  13.78 $\pm$  0.06  &    -2.17  $\pm$  0.37     &  -1.61  $\pm$    0.42   \\
& 3.034429$\dagger$  &    15 $\pm$   6  &  13.71  $\pm$  0.158  &   13.79  $\pm$  0.36  &  14.53 $\pm$  0.12  &    -1.86  $\pm$  0.08     &  -1.59  $\pm$    0.13   \\
& 3.034765$\dagger$  &    19 $\pm$   5  &  13.28  $\pm$  0.218  &   13.69  $\pm$  0.17  &  14.65 $\pm$  0.08  &    -2.05  $\pm$  0.11     &  -2.02  $\pm$    0.15   \\
& 3.035330           &    21 $\pm$  20  &  12.79  $\pm$  0.421  &   12.85  $\pm$  0.29  &  13.61 $\pm$  0.06  &    -2.01  $\pm$  0.19     &  -1.64  $\pm$    0.37   \\
& 3.035959           &    16 $\pm$  12  &  13.51  $\pm$  0.091  &   12.94  $\pm$  0.20  &  14.72 $\pm$  0.09  &    -1.78  $\pm$  0.06     &  -1.88  $\pm$    0.08   \\
& 3.036588           &    16 $\pm$  11  &  13.59  $\pm$  0.133  &   13.80  $\pm$  0.15  &  15.35 $\pm$  0.40  &    -1.92  $\pm$  0.10     &  -2.10  $\pm$    0.20   \\
& 3.036888           &    13 $\pm$   3  &  13.64  $\pm$  0.160  &   14.70  $\pm$  0.39  &  14.31 $\pm$  0.35  &    -2.01  $\pm$  0.30     &  -1.48  $\pm$    0.28   \\
& 3.037293           &    15 $\pm$   3  &  13.23  $\pm$  0.206  &   12.50  $\pm$  0.50  &  14.54 $\pm$  0.05  &    -1.75  $\pm$  0.15     &  -2.01  $\pm$    0.17   \\

\hline
\Jzuts

& 2.777806$\dagger$   &  15  $\pm$  7  &  12.94 $\pm$   0.13  &  13.00  $\pm$  0.07  &  13.35  $\pm$  0.13  &  -2.07  $\pm$   0.05    &      -1.28  $\pm$    0.12   \\ 
& 2.778885$\dagger$   &  51  $\pm$  7  &  13.43 $\pm$   0.08  &  13.70  $\pm$  0.06  &  14.16  $\pm$  0.04  &  -2.13  $\pm$   0.04    &      -1.57  $\pm$    0.07   \\ 
& 2.779292$\dagger$   &   9  $\pm$ 13  &  13.16 $\pm$   0.12  &  14.09  $\pm$  0.68  &  14.55  $\pm$  0.87  &  -2.39  $\pm$   0.15    &      -2.16  $\pm$    0.20   \\ 
& 2.779424$\dagger$   &  43  $\pm$  6  &  11.90 $\pm$   0.55  &  13.44  $\pm$  0.72  &  14.51  $\pm$  0.16  &  -2.80  $\pm$   0.35    &      -2.94  $\pm$    0.45   \\ 
& 2.779803$\dagger$   &  12  $\pm$  6  &  13.04 $\pm$   0.12  &  13.53  $\pm$  0.20  &  13.66  $\pm$  0.41  &  -2.24  $\pm$   0.06    &      -1.47  $\pm$    0.12   \\ 
& 2.783816            &  33  $\pm$  2  &  13.46 $\pm$   0.04  &  <12.27              &  14.05  $\pm$  0.01  &  >-1.60                 &      -1.44  $\pm$    0.04   \\

\hline
\Judts

& 3.025500            &  24  $\pm$   7  &    12.80 $\pm$ 0.24  &   12.91  $\pm$  0.15 &  14.04  $\pm$  0.04 &   -2.09   $\pm$   0.10     &      -2.03    $\pm$  0.24     \\
& 3.026000            &  24  $\pm$  29  &    13.01 $\pm$ 0.22  &   12.50  $\pm$  0.37 &  13.94  $\pm$  0.05 &   -1.85   $\pm$   0.08     &      -1.75    $\pm$  0.20     \\
& 3.026410            &  7  $\pm$    5  &    13.10 $\pm$ 0.21  &   13.43  $\pm$  0.28 &  14.53  $\pm$  0.26 &   -2.17   $\pm$   0.10     &      -2.20    $\pm$  0.21     \\
& 3.026770            &  6  $\pm$   18  &    11.42 $\pm$ 1.25  &   12.62  $\pm$  0.29 &  14.02  $\pm$  0.28 &   -2.63   $\pm$   0.60     &      -3.15    $\pm$  0.65     \\
& 3.027100            &  19  $\pm$  10  &    12.84 $\pm$ 0.34  &   12.84  $\pm$  0.15 &  13.82  $\pm$  0.04 &   -2.04   $\pm$   0.15     &      -1.79    $\pm$  0.31     \\
& 3.027610            &  4  $\pm$    2  &    12.72 $\pm$ 0.48  &   12.35  $\pm$  0.46 &  13.63  $\pm$  0.24 &   -1.90   $\pm$   0.19     &      -1.74    $\pm$  0.45     \\
& 3.027930$\dagger$   &  11  $\pm$   9  &    11.81 $\pm$ 0.76  &   12.99  $\pm$  0.20 &  13.92  $\pm$  0.09 &   -2.62   $\pm$   0.40     &      -2.97    $\pm$  0.50     \\
& 3.028250$\dagger$   &  20  $\pm$   5  &    13.04 $\pm$ 0.20  &   13.51  $\pm$  0.07 &  14.36  $\pm$  0.04 &   -2.24   $\pm$   0.11     &      -2.11    $\pm$  0.20     \\
& 3.029030$\dagger$   &  17  $\pm$   4  &    12.76 $\pm$ 0.33  &   13.45  $\pm$  0.08 &  14.33  $\pm$  0.04 &   -2.34   $\pm$   0.16     &      -2.35    $\pm$  0.30     \\
& 3.029300$\dagger$   &  22  $\pm$  13  &    13.06 $\pm$ 0.19  &   12.65  $\pm$  0.44 &  13.74  $\pm$  0.06 &   -1.88   $\pm$   0.07     &      -1.52    $\pm$  0.17     \\
& 3.030400$\dagger$   &  19  $\pm$  21  &    13.35 $\pm$ 0.09  &   12.69  $\pm$  0.42 &  13.32  $\pm$  0.22 &   -1.79   $\pm$   0.03     &      -0.74    $\pm$  0.16     \\
& 3.030780$\dagger$   &  16  $\pm$   6  &    14.41 $\pm$ 0.05  &   13.87  $\pm$  0.11 &  15.49  $\pm$  0.24 &   -1.79   $\pm$   0.07     &      -1.80    $\pm$  0.05     \\
& 3.031220$\dagger$   &  10  $\pm$   6  &    13.77 $\pm$ 0.15  &   13.28  $\pm$  0.15 &  14.44  $\pm$  1.15 &   -1.85   $\pm$   0.06     &      -1.51    $\pm$  0.15     \\
& 3.031600$\dagger$   &  11  $\pm$   8  &    15.74 $\pm$ 0.47  &   13.95  $\pm$  0.31 &  14.57  $\pm$  1.29 &   -1.13   $\pm$   0.40     &      ---                      \\
& 3.031700$\dagger$   &  16  $\pm$  30  &    12.67 $\pm$ 0.75  &   13.33  $\pm$  0.31 &  14.69  $\pm$  0.87 &   -2.33   $\pm$   0.34     &      -2.76    $\pm$  0.20     \\
& 3.032300$\dagger$   &  13  $\pm$   8  &    13.94 $\pm$ 0.09  &   13.54  $\pm$  0.16 &  14.37  $\pm$  1.45 &   -1.87   $\pm$   0.04     &      -1.29    $\pm$  0.20     \\
& 3.032600$\dagger$   &  15  $\pm$   5  &    14.02 $\pm$ 0.07  &   14.20  $\pm$  0.18 &  15.43  $\pm$  0.74 &   -2.04   $\pm$   0.04     &      -2.07    $\pm$  0.22     \\
& 3.033230$\dagger$   &  27  $\pm$  14  &    14.40 $\pm$ 0.02  &   13.42  $\pm$  0.18 &  14.99  $\pm$  1.10 &   -1.65   $\pm$   0.02     &      -1.42    $\pm$  0.20     \\
& 3.033730            &  9  $\pm$    3  &    13.99 $\pm$ 0.08  &   12.10  $\pm$  0.73 &  15.19  $\pm$  0.55 &   -1.35   $\pm$   0.10     &      -1.95    $\pm$  0.16     \\
& 3.034170            &  11  $\pm$   4  &    13.28 $\pm$ 0.08  &   <12.25             &  13.98  $\pm$  0.07 &   >-1.66                   &      -1.54    $\pm$  0.07     \\

\hline
\Judqh
& 2.523914           &  10 $\pm$   1  &    13.82 $\pm$   0.21 &  12.23 $\pm$  0.91 &   14.12 $\pm$   0.30  &  -1.45  $\pm$  0.30  &   -1.16 $\pm$     0.20 \\

& 2.527214           & 27 $\pm$    1  &    12.84 $\pm$   0.10 &  13.30 $\pm$  0.01 &   14.53 $\pm$   0.01  &  -2.23  $\pm$  0.05  &   -2.45 $\pm$     0.09 \\
& 2.527698$\dagger$  &  4 $\pm$    3  &    12.37 $\pm$   0.26 &  12.41 $\pm$  0.28 &   13.20 $\pm$   0.30  &  -2.06  $\pm$  0.11  &   -1.64   $\pm$    0.26 \\
& 2.528276$\dagger$  & 34 $\pm$    4  &    13.07 $\pm$   0.07 &  13.07 $\pm$  0.05 &   14.09 $\pm$   0.03  &  -2.04  $\pm$  0.03  &   -1.83 $\pm$     0.06 \\
& 2.528914$\dagger$  & 14 $\pm$    3  &    12.84 $\pm$   0.10 &  13.22 $\pm$  0.03 &   13.90 $\pm$   0.04  &  -2.20  $\pm$  0.05  &   -1.86 $\pm$     0.10 \\
& 2.529348$\dagger$  &  9 $\pm$    5  &    12.88 $\pm$   0.18 &  13.11 $\pm$  0.19 &   13.65 $\pm$   0.48  &  -2.14  $\pm$  0.08  &   -1.60 $\pm$     0.16 \\
& 2.529771$\dagger$  & 25 $\pm$    8  &    13.08 $\pm$   0.21 &  13.56 $\pm$  0.11 &   14.59 $\pm$   0.07  &  -2.23  $\pm$  0.10  &   -2.27 $\pm$     0.20 \\
& 2.530486$\dagger$  & 38 $\pm$    3  &    13.83 $\pm$   0.02 &  13.82 $\pm$  0.03 &   14.78 $\pm$   0.03  &  -2.00  $\pm$  0.02  &   -1.75 $\pm$     0.03 \\

\hline
\Juttu

& 2.885911            &  31  $\pm$  5   &    13.76 $\pm$  0.10   &   12.44  $\pm$ 0.27    &  13.42 $\pm$   0.09  &  -1.53   $\pm$   0.05    &      -0.21  $\pm$      0.18  \\
& 2.886689            &  3   $\pm$  2   &    13.16 $\pm$  0.54   &   12.40  $\pm$ 0.63    &  13.53 $\pm$   0.37  &  -1.74   $\pm$   0.20    &      -1.23  $\pm$      0.50  \\  
& 2.887006            &  7   $\pm$  2   &    13.78 $\pm$  0.27   &   12.61  $\pm$ 0.27    &  13.83 $\pm$   0.12  &  -1.57   $\pm$   0.10    &      -0.83  $\pm$      0.33  \\
& 2.890579            &  6   $\pm$  5   &    13.51 $\pm$  0.56   &   <12.04               &  13.28 $\pm$   0.43  &  >-1.49                  &       -0.41 $\pm$  0.23  \\
& 2.890883            &  5   $\pm$  8   &    13.94 $\pm$  0.94   &   <12.04               &  14.54 $\pm$   0.41  &  >-1.32                  &      -1.37  $\pm$ 0.65  \\
& 2.891287            &  20  $\pm$  30  &    14.35 $\pm$  0.31   &   <12.04               &  13.91 $\pm$   0.06  &  >-1.18                  &       0.01  $\pm$  0.35  \\
& 2.891717            &  15  $\pm$  10  &    13.79 $\pm$  0.72   &   <12.04               &  13.70 $\pm$   0.09  &  >-1.39                  &      -0.63  $\pm$ 0.45  \\
& 2.892126            &  5   $\pm$  1   &    12.83 $\pm$  0.82   &   <12.04               &  13.31 $\pm$   0.64  &  >-1.73                  &      -1.32   $\pm$ 0.43  \\

\hline
\Jutch

& 2.877874          &  21  $\pm$     9  &     13.30  $\pm$   0.08  &  <11.83              &  <11.58             &  >-1.50                     &      ---                    \\
& 2.878764          &  11  $\pm$    13  &     13.06  $\pm$   0.75  &  <11.83              &  12.83 $\pm$  0.10  &  >-1.58                     &      -0.40   $\pm$    0.17  \\
& 2.879557          &  78  $\pm$     4  &     14.01  $\pm$   0.02  &  <11.83              &  13.21 $\pm$  0.08  &  >-1.25                     &      +0.75   $\pm$    0.15  \\
& 2.880677          &  40  $\pm$     3  &     13.82  $\pm$   0.02  &  <11.83              &  13.44 $\pm$  0.03  &  >-1.31                     &      -0.16   $\pm$    0.09  \\
& 2.881652          &   4  $\pm$     1  &     13.67  $\pm$   0.10  &  <11.83              &  13.18 $\pm$  0.25  &  >-1.37                     &      +0.05   $\pm$    0.20  \\
& 2.883319          &  67  $\pm$     2  &     14.55  $\pm$   0.01  &  <11.83              &  14.16 $\pm$  0.02  &  >-1.05                     &      -0.12   $\pm$    0.04  \\ 
& 2.884272          &  16  $\pm$     1  &     15.04  $\pm$   0.07  &  12.60  $\pm$  0.07  &  13.70 $\pm$  0.15  &   -1.13   $\pm$   0.07      &      0.81    $\pm$    0.15  \\
& 2.884877          &  25  $\pm$     1  &     15.68  $\pm$   0.25  &  12.99  $\pm$  0.07  &  14.93 $\pm$  0.16  &   -0.74   $\pm$   0.30      &      ---                    \\
& 2.885429          &   7  $\pm$     1  &     13.27  $\pm$   0.18  &  <11.83              &  14.52 $\pm$  0.33  &  >-1.51                     &     -2.02    $\pm$ 0.18     \\
& 2.885972          &   8  $\pm$     1  &     14.32  $\pm$   0.08  &  12.31  $\pm$  0.14  &  <11.58             &   -1.30   $\pm$   0.08      &      ---                    \\ 
& 2.886478          &  14  $\pm$     4  &     14.81  $\pm$   0.23  &  12.78  $\pm$  0.07  &  14.37 $\pm$  0.28  &   -1.27   $\pm$   0.15      &      -0.03   $\pm$  0.35    \\
& 2.886944          &  55  $\pm$     3  &     14.21  $\pm$   0.05  &  <12.21              &  13.72 $\pm$  0.07  &  >-1.31                     &     0.02     $\pm$  0.04    \\
& 2.887575          &  21  $\pm$    11  &     12.82  $\pm$   0.49  &  <12.21              &  <11.58             &  >-1.81                     &       ---                   \\
& 2.887927          &   4  $\pm$     1  &     14.66  $\pm$   0.31  &  <12.21              &  13.23 $\pm$  0.05  &  >-1.14                     &       1.03   $\pm$    0.25  \\
& 2.888483          &   2  $\pm$     1  &     15.73  $\pm$   0.41  &  12.64  $\pm$  0.21  &  13.53 $\pm$  0.03  &   -0.61   $\pm$   0.60      &      ---                    \\
& 2.889241          &  49  $\pm$    16  &     14.11  $\pm$   0.01  &  <12.20              &  13.37 $\pm$  0.04  &  >-1.33                     &       0.56   $\pm$    0.10  \\
& 2.889963          &   5  $\pm$     8  &     14.97  $\pm$   0.13  &  <12.20              &  13.68 $\pm$  0.03  &  >-1.02                     &        0.86  $\pm$     0.22 \\
& 2.890438          &   8  $\pm$     1  &     14.75  $\pm$   0.09  &  <12.20              &  13.88 $\pm$  0.02  &  >-1.11                     &        0.77  $\pm$     0.07 \\
& 2.890932          &  10  $\pm$     1  &     14.30  $\pm$   0.04  &  <12.20              &  13.31 $\pm$  0.08  &  >-1.27                     &       0.95   $\pm$    0.06  \\
& 2.891472          &   5  $\pm$     1  &     14.42  $\pm$   0.11  &  <12.20              &  13.70 $\pm$  0.05  &  >-1.22                     &       0.78   $\pm$    0.08  \\
& 2.891999$\dagger$ &  17  $\pm$     5  &     13.84  $\pm$   0.11  &  14.08  $\pm$  0.13  &  14.89 $\pm$  0.13  &   -1.98   $\pm$   0.08      &    -1.81     $\pm$ 0.20     \\
& 2.892758$\dagger$ &  21  $\pm$     1  &     13.74  $\pm$   0.06  &  14.63  $\pm$  0.07  &  <11.58             &   -2.13   $\pm$   0.04      &      ---                    \\
& 2.893164$\dagger$ &  35  $\pm$     1  &     14.38  $\pm$   0.01  &  13.92  $\pm$  0.04  &  15.66 $\pm$  0.05  &   -1.71   $\pm$   0.02      &      -1.75   $\pm$  0.04    \\
& 2.894011$\dagger$ &  22  $\pm$     3  &     13.05  $\pm$   0.38  &  13.63  $\pm$  0.12  &  14.42 $\pm$  0.09  &   -2.23   $\pm$ 0.16        &     -2.14    $\pm$ 0.36     \\
& 2.894848          &  66  $\pm$     1  &     14.23  $\pm$   0.01  &  13.71  $\pm$  0.02  &  14.70 $\pm$  0.05  &   -1.75   $\pm$  0.02       &      -1.31   $\pm$  0.04    \\
& 2.896053          &  39  $\pm$     8  &     13.63  $\pm$   0.05  &  <11.87              &  <11.58             &   >-1.39                    &      ---                    \\

\hline
\Jdddh
& 2.768395           &  11 $\pm$   1 &   14.49 $\pm$  0.03  &  13.23  $\pm$  0.03 &   15.49 $\pm$  0.12  &  -1.56  $\pm$    0.02  & -1.73  $\pm$     0.04  \\
& 2.768918$\dagger$  &  10 $\pm$   1 &   14.15 $\pm$  0.04  &  14.23  $\pm$  0.14 &   15.40 $\pm$  0.67  &  -1.99  $\pm$    0.02  & -1.96  $\pm$     0.15  \\
& 2.769360$\dagger$  &   9 $\pm$   1 &   14.13 $\pm$  0.03  &  14.44  $\pm$  0.13 &   14.79 $\pm$  0.15  &  -2.04  $\pm$    0.03  & -1.49  $\pm$     0.07  \\
& 2.769923$\dagger$  &  29 $\pm$   2 &   12.82 $\pm$  0.12  &  12.87  $\pm$  0.04 &   13.64 $\pm$  0.08  &  -2.07  $\pm$    0.05  & -1.65  $\pm$     0.12  \\

\label{t:column_density}
\end{longtable}
\tablefoot{$\dagger$ Component within the $\Delta v_{90}$ range of low-ionization metals.}
\twocolumn

\clearpage

\section{Effect of the metallicity in the models\label{s:metals}}

\begin{figure}[h!]
    \centering
    \includegraphics[width = 0.9\hsize,trim={0.2cm 0 1.5cm 0.8cm},clip]{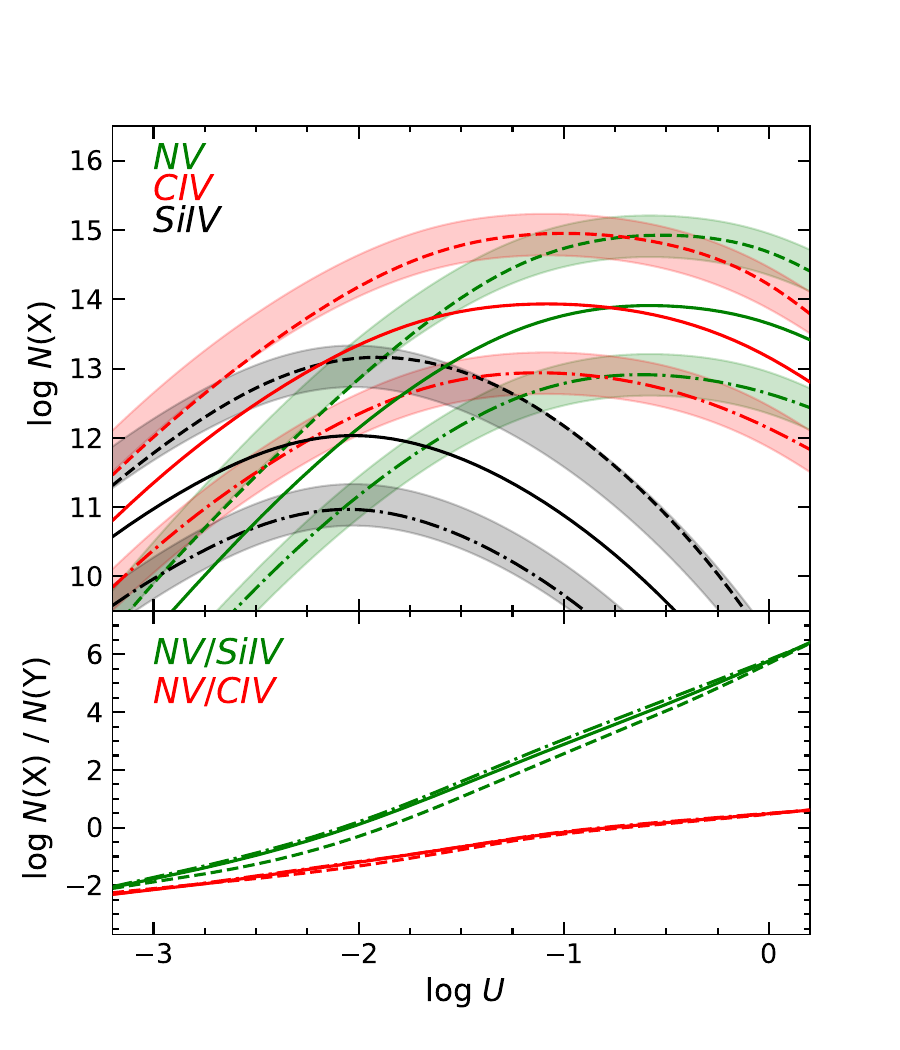}
    \caption{Predicted column density of \NV, \SiIV\ and \CIV\ (top) and \NV-to-\SiIV\ and \NV-to-\CIV\ ratio (bottom) as a function of the ionization parameter. The solid lines correspond to the $\log N(\HI)=15$ model in Fig.~\ref{f:N_vs_U}. The dashed (resp. dashed-dotted) lines correspond to models assuming a 10 times higher (resp. lower) metallicity. 
    The shaded areas shown results from the central model (solid line), simply scaled up and down {a posteriori} by 1~dex. These scaled results match the predictions from the full calculations with corresponding metallicities reasonably well, typically within a factor of two (illustrated by the width of the stripes). }
    \label{f:N_and_ratio_N_vs_U_dif_Z}
\end{figure}
\end{appendix}

\addtolength{\tabcolsep}{-3pt}
\addtolength{\tabcolsep}{3pt}

\clearpage
\clearpage

\end{document}